\newif{\iffinal}                             % camera-ready copy (or review final)
\newif{\ifreviews}
\newcounter{numauthors}                      % number of defined authors
\newcommand{\newauthor}[4]{                  % define author fields
    \expandafter\def\csname name:#1\endcsname{#2}
    \expandafter\def\csname email:#1\endcsname{#3}
    \expandafter\def\csname inst:#1\endcsname{#4}
    \stepcounter{numauthors}}
\newif{\iflinks}

\finaltrue{}                                 % hide or show authors
\reviewstrue
\linksfalse{}

\newcommand{\sblock}[1]{

\vspace{3pt}
\noindent{\bf #1}
}

\newcommand{\refappendix}[1]{\hyperref[#1]{Appendix~\ref*{#1}}}

\newcommand{\papertitle}{Measuring and Mitigating the \\ Risk of IP Reuse on Public Clouds}

\makeatletter
\newcommand{\linebreakand}{%
  \end{@IEEEauthorhalign}
  \hfill\mbox{}\par
  \mbox{}\hfill\begin{@IEEEauthorhalign}
}
\makeatother

\newcommand{\authors}{}                      % holds author list
\makeatletter\newcommand{\addauthor}[1]{%    % pre-build author list
    \g@addto@macro\authors{#1}}\makeatother  
\newcommand{\showname}[1]{\expandafter\csname name:#1\endcsname}
\newcommand{\showemail}[1]{\expandafter\csname email:#1\endcsname}
\newcommand{\showinst}[1]{\expandafter\csname inst:#1\endcsname}
\newcommand\auth[2]{%
    \if0#2
        \showname{#1}\\%
        \showemail{#1}\\%
        \showinst{#1}%
    \fi
    \if1#2
        \showname{#1}%
    \fi
    \if2#2
        \showemail{#1}%
    \fi
    \if3#2
        \showinst{#1}%
    \fi}

\documentclass[conference]{IEEEtran}
\newcommand\buildauthors{
    \newcount\currentauthor
    \currentauthor=1
    \loop\ifnum\currentauthor<\numexpr\value{numauthors}+1
        \expandafter\addauthor\expandafter{%
            \expandafter\IEEEauthorblockN\expandafter{%
                \expandafter\auth\the\currentauthor{0}}\and
        }
        \advance\currentauthor+1
    \repeat
}\buildauthors
\author{\authors}
\title{\papertitle}
\usepackage[ruled,linesnumbered]{algorithm2e}    % Floating algorithm environment with algorithmic keywords
\usepackage{amsmath}                             % AMS mathematical facilities
\usepackage{amsthm}                              % Typesetting theorems (AMS style)
\usepackage{booktabs}                            % Publication quality tables in LaTeX
\usepackage{centernot}                           % Centered \not command
\usepackage{colortbl}                            % Add colour to LaTeX tables
\usepackage{graphicx}                            % Enhanced support for graphics
\usepackage{hhline}                              % Better horizontal lines in tabulars and arrays
\usepackage{lipsum}                              % Easy access to the Lorem Ipsum dummy text
\usepackage{mathtools}                           % Mathematical tools to use with amsmath
\usepackage[all]{nowidow}                        % Avoid widows
\usepackage{optidef}                             % Environments for writing optimization problems
\usepackage{pgfplotstable}                       % Loads and processes numerical tables
\usepackage{subcaption}                          % Support for sub-captions
\usepackage{siunitx}                             % International System of Units
\usepackage{substr}                              % Substring Functions with LaTeX
\usepackage{tikz}                                % Create PostScript and PDF graphics in TEX
\usepackage{tcolorbox}                           % Coloured boxes, for LATEX examples and theorems, etc
\usepackage{xcolor}                              % Driver-independent color extensions for LaTeX and pdfLaTeX
\usepackage[breaklinks]{hyperref}                           % Create hyperlinks in PDF
\usepackage{xcolor}                              % Text Coloring
\usepackage{xspace}                              % Text Coloring
\usepackage{multirow}
\usepackage{float}
\usepackage{url}
\usepackage{breakurl}
\usepackage{pgf}
\usepackage{wrapfig}
\usepackage{fancyhdr}

\usepackage{pifont}% http://ctan.org/pkg/pifont   % for digits in black circle

\newenvironment{sidebar}                            % simple sidebar 
    {\figure[!t]
    \tcolorbox}
    {\endtcolorbox
    \endfigure}

\hypersetup{colorlinks, linkcolor=blue}
\usetikzlibrary{calc, fit, shapes.geometric}
\pgfplotsset{layers/layerss/.define layer set={% % setup layers for tikz pictures
    bg,main,fg}{},set layers=layerss,}
\SetArgSty{textup}                               % do not force italics in algorithm conditionals
\pgfplotsset{compat=1.16,%                       % use stable package version
    compat/show suggested version=false}         % disable warnings and messages
\DeclareSIUnit[number-unit-product=]\percent{%   % remove added space when using %
    \char`\%}
\newcommand{\ie}{i.e.\xspace}                    % id est
\newcommand{\eg}{e.g.\xspace}                    % exempli gratia
\begin{document}

\fancypagestyle{firststyle}
{
   \fancyhf{}
   \chead{\textit{--------------------------------------- In 43rd IEEE Symposium on Security and Privacy ---------------------------------------}}
   \renewcommand{\headrulewidth}{0pt} % removes horizontal header line
}
\pagenumbering{gobble}
\author{\\\IEEEauthorblockN{Eric Pauley}
\IEEEauthorblockA{%
\textit{Computer Science \& Engineering}\\
\textit{The Pennsylvania State University}\\
epauley@psu.edu}
\and
\\\IEEEauthorblockN{Ryan Sheatsley}
\IEEEauthorblockA{%
\textit{Computer Science \& Engineering}\\
\textit{The Pennsylvania State University}\\
sheatsley@psu.edu}
\and
\\\IEEEauthorblockN{Blaine Hoak}
\IEEEauthorblockA{%
\textit{Computer Science \& Engineering}\\
\textit{The Pennsylvania State University}\\
bhoak@psu.edu}
\linebreakand
\IEEEauthorblockN{Quinn Burke}
\IEEEauthorblockA{%
\textit{Computer Science \& Engineering}\\
\textit{The Pennsylvania State University}\\
qkb5007@psu.edu}
\and
\IEEEauthorblockN{Yohan Beugin}
\IEEEauthorblockA{%
\textit{Computer Science \& Engineering}\\
\textit{The Pennsylvania State University}\\
ybeugin@psu.edu}
\and
\IEEEauthorblockN{Patrick McDaniel}
\IEEEauthorblockA{%
\textit{Computer Science \& Engineering}\\
\textit{The Pennsylvania State University}\\
mcdaniel@cse.psu.edu}
}
\maketitle
\thispagestyle{firststyle}
\pagestyle{plain}
\noindent\begin{abstract}\label{abstract}
Public clouds provide scalable and cost-efficient computing through resource sharing.
However, moving from traditional on-premises service management to clouds introduces new challenges; failure to correctly provision, maintain, or decommission elastic services can lead to functional failure and vulnerability to attack.
In this paper, we explore a broad class of attacks on clouds which we refer to as cloud squatting. In a cloud squatting attack, an adversary allocates resources in the cloud (\eg, IP addresses) and thereafter leverages latent configuration to exploit prior tenants.
To measure and categorize cloud squatting we deployed a custom Internet telescope within the Amazon Web Services \texttt{us-east-1} region.
Using this apparatus, we deployed over \SI{3}{} million servers receiving \SI{1.5}{} million unique IP addresses ($\approx \SI{56}{\percent}$ of the available pool) over \SI{101}{} days beginning in March of 2021. We identified \SI{4}{} classes of cloud services, \SI{7}{} classes of third-party services, and DNS as sources of exploitable latent configurations. We discovered that exploitable configurations were both common and in many cases extremely dangerous; we received over \SI{5}{} million cloud messages, many containing sensitive data such as financial transactions, GPS location, and PII. Within the \SI{7}{} classes of third-party services, we identified dozens of exploitable software systems spanning hundreds of servers (e.g., databases, caches, mobile applications, and web services). Lastly, we identified \SI{5446}{} exploitable domains spanning \SI{231}{} eTLDs—including \SI{105}{} in the top \SI{10000}{} and \SI{23}{} in the top \SI{1000}{} popular domains. 
Through tenant disclosures we have identified several root causes, including (a) a lack of organizational controls, (b) poor service hygiene, and (c) failure to follow best practices.
We conclude with a discussion of the space of possible mitigations and describe the mitigations to be deployed by Amazon in response to this study.
\end{abstract}

\setlength{\textfloatsep}{10pt plus 0pt minus 5pt}
\section{Introduction}
Public clouds such as Amazon Web Services~\cite{aws_website}, Google Cloud~\cite{google_cloud_website}, and Microsoft Azure~\cite{azure_website} offer a myriad of benefits to tenants; by providing a virtual private data center on top of shared infrastructure, clouds allow users to scale services with changing demand, recover from faults, and fluidly and autonomously provision services. As such, public clouds are now used by almost every major computing enterprise. At the same time, the sharing of resources of clouds offers unique architectural and security challenges. A hard-learned truism is that care must be taken during provisioning, use, and decommissioning of cloud servers. 

\begin{figure}[t]
\centering
\includegraphics[width=0.7\columnwidth]{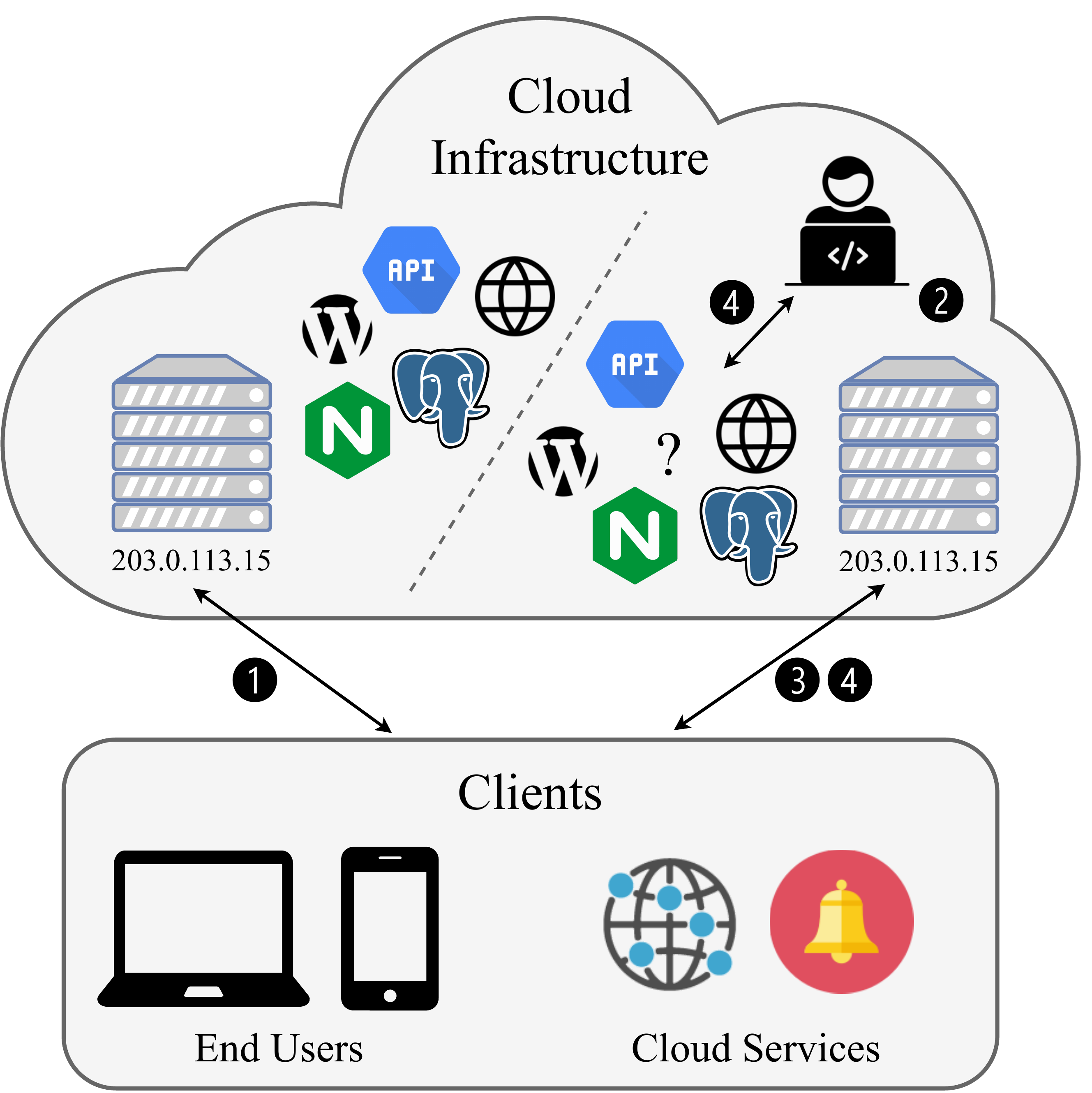}
 \caption{Exploiting IP Reuse---\ding{202} A client is configured to connect to a tenant's service. \ding{203} After the service is retired, an adversary provisions a server and is granted an IP address previously assigned to the tenant. \ding{204} The client connects to the adversary-controlled server. \ding{205} The adversary receives traffic intended for the tenant, infers the intended service from the client traffic, and exploits it. Note that \texttt{203.0.113.15} is reserved for example usage by IANA.}
\label{fig:overview}
\end{figure}

Clouds providing elastic computing assign IP addresses (or IPs) from a shared pool. An IP address can be used by a deployed server instance and referenced by managed cloud services (e.g., server-to-server messaging), third-party applications deployed by the tenant (e.g., using the deployed server as a database or computer resource), or directly configured in Domain Name System (DNS) records. Importantly, when the service is decommissioned, the IP is released and may be reallocated to other tenants~\cite{borgolte_cloud_2018}. This decommissioning process and shared IP architecture causes a vulnerability: configurations (e.g., dangling DNS~\cite{liu_all_2016}) that are not updated in tandem with the decommissioning process will continue to refer to the (now obsolete) IP. We refer to these configurations as \textit{latent configurations}. If an adversary subsequently acquires the IP (\autoref{fig:overview}), they can exploit the latent configuration by receiving traffic intended for the previous service or masquerading as the prior tenant. While previous studies have measured the risk of latent configuration and IP use-after-free through DNS, the community currently lacks understanding of how other types of configuration could lead to vulnerabilities. In this work, we investigate vulnerabilities caused by latent configuration broadly, namely cloud service configuration and third-party service configuration, in addition to dangling DNS. We name the resulting superset class of attacks \textit{cloud squatting}. A cloud squatting attack occurs when an adversary acquires some reused cloud resource (e.g., an IP address) that is referenced by latent configuration (e.g., cloud service configuration).

We hypothesize that, not only does latent configuration span beyond DNS, but that these previously unexplored vulnerabilities are widespread and readily discoverable. We characterize three classes of latent configuration in clouds: (1) configuration in managed cloud services, (2) configuration in software deployed by tenants, and (3) configuration through DNS. Each of these classes (shown in \autoref{tbl:vulnerability_taxonomy}) presents distinct challenges and opportunities for study as we evaluate \textit{legitimacy} (\ie, actual traffic intended for a previous tenant) and \textit{exploitability} (\ie, ability of an adversary to receive sensitive data).

To measure the presence of latent configuration in clouds, we developed and deployed an Internet telescope in the Amazon Web Services \texttt{us-east-1} region\footnote{We selected the AWS \texttt{us-east-1} region because of its size and diversity. We expect similar results in any public cloud, as many root causes are unrelated to AWS or its services.}. Using this telescope, we provisioned over \SI{3}{} million cloud servers with \SI{1.5}{} million unique IPs ($\approx$ \SI{56}{\percent} of the region IP address pool). We passively collected all inbound traffic (containing \SI{596}{M} TCP sessions) over \SI{101}{} days starting on March 8, 2021. The scale of this experiment and the size of the cloud studied ensure a representative view of traffic in a commercial public cloud. 

\begin{table*}[h]
\centering
\caption{Taxonomy of discovered cloud squatting vulnerabilities.}
\begin{tabular}{lp{0.42\linewidth}p{0.35\linewidth}}
\toprule
\multicolumn{1}{c}{\textbf{Vulnerability Class}} & \textbf{Description} & \textbf{Characterization}\\
\midrule
\multicolumn{3}{l}{\underline{\textit{Cloud Configuration}}}\\
\multicolumn{1}{r}{SNS~\cite{sns}} & Topics send messages to IP addresses reallocated to another tenant.& \SI{24.9}{k} IPs received \SI{1.6}{M} messages \\
\multicolumn{1}{r}{Route53~\cite{route53}} & Service health checks are sent to previously-controlled IP addresses. An adversary who responds correctly to these health checks can receive live traffic. & \SI{2.8}{k} IPs received \SI{3.6}{M} messages\\
\multicolumn{1}{r}{CloudFront~\cite{cloudfront}} & Web requests are automatically forwarded to raw IPs that a tenant may no longer control. & \SI{65}{} IPs received \SI{1.7}{k} messages\\
\multicolumn{1}{r}{API Gateway~\cite{api_gateway}} & Web requests are forwarded to previously-configured backends that a tenant no longer controls. & 3 IPs received 10 messages\\
&\\
\multicolumn{3}{l}{\underline{\textit{Third-Party Services}}}\\
\multicolumn{1}{r}{Databases} & {Database traffic with sensitive information.}& Postgres~\cite{postgres}, ElasticSearch~\cite{elasticsearch}, MySQL~\cite{mysql}, etc.\\
\multicolumn{1}{r}{Caches} & {Cache traffic that could allow gaining control of caches. }& {Redis~\cite{redis} Cluster traffic}\\
\multicolumn{1}{r}{Financial} & {Traffic containing credentials and potentially transaction data. }& {Financial Information eXchange (FIX)~\cite{fix_protocol}} \\
\multicolumn{1}{r}{Logging} & {Crash reports/logs from smartphone apps. } & Tracebacks, emails, and device identifiers\\
\multicolumn{1}{r}{Metrics} & {Analytics on user actions from smartphone and web applications. }& PII and website actions\\
\multicolumn{1}{r}{Webhooks} & {Applications send events intended for another tenant. }& {Teamwork~\cite{teamwork}, BitBucket~\cite{bitbucket}, Segment~\cite{segment}, etc.}\\
\multicolumn{1}{r}{App Traffic} & {Smartphone/web applications connecting to APIs via raw IPs. }& Private info received from apps \\
&&\\
\underline{{\textit{Domain Name Resolution}}} & DNS entries pointing to IP addresses that tenants no longer control. & \SI{5446}{} domains \\
\bottomrule
\end{tabular}
\label{tbl:vulnerability_taxonomy}
\end{table*}

The experiment overwhelmingly confirmed our hypothesis and demonstrated a surprising prevalence of latent configuration. We received over \SI{5}{} million messages directed to prior tenants on over \SI{27}{} thousand IPs that we were assigned from cloud-managed services (e.g., SNS messaging). These messages contained sensitive data including financial transaction metadata, customer GPS location history, and customer PII (e.g., driver’s license data and personal addresses). Third-party services likewise exposed sensitive data, with hundreds of instances across 7 classes of services found vulnerable. With respect to latent DNS configurations, we identified \SI{5446}{} (second level, \eg, \texttt{example.com}) domains spanning \SI{231}{} eTLDs (\eg, \texttt{.com})—including \SI{105}{} in the top \SI{10000}{} and \SI{23}{} in the top \SI{1000}{} list of popular domains. Moreover, the results were observed across the entirety of cloud tenant populations: in government, academic, and industrial (e.g., high-tech, financial, health care, and entertainment) organizations. Following our initial disclosure, Amazon performed an internal review of customer configurations which found latent configurations in similar breadth and quantity in all of their regions.

Summarized in \autoref{tbl:vulnerability_taxonomy}, we identified traffic sourced from \SI{4}{} kinds of cloud services, \SI{7}{} classes of third-party services, and DNS as sources of exploitable latent configurations. Latent configurations of cloud services supporting messaging, health checks, content delivery and generalized API controls were observed. We also found that third-party services can produce latent configurations as diverse as the applications they support. Here, we found cases of latent configurations in databases, distributed caches, logging, and many others. Lastly, we found many cases of poorly managed DNS leading to exploitable organizations.

We contacted Amazon AWS in June of 2021 and have since worked with them to support coordinated disclosures and develop mitigations. We are also working with US government agencies and other cloud providers to support detection and disclosure (see \refappendix{appendix:disclosure}). We conducted virtual meetings with select tenants for disclosure and to discuss root causes. Broadly, the root causes identified include, (a) a lack of organizational control over cloud accounts, (b) poor service hygiene (e.g., poor or uncontrolled management of service configurations), and (c) failure of engineers/departments to follow organizational policies and best practices, and (d) incomplete automation. Both our measurement study and tenant disclosures were covered by an IRB exemption from our host institution, and followed ethical considerations consistent with contemporary works in network measurement.

Lastly, we have developed and evaluated a set of mitigations to prevent latent configuration vulnerabilities and reduce an adversary's ability to acquire IPs associated with vulnerable tenants. Existing best practices, such as cloud configuration auditing tools, reserved IP blocks, and managed configuration can prevent latent configuration when properly used by tenants. However, even when tenants are unable to adopt new best practices, changes to the IP allocation pool can prevent adversaries from successfully carrying out a cloud squatting attack. We introduce such a technique, which we name \textit{IP Tagging}, reducing adversaries' access to tenant IPs by \SI{99.94}{\percent} over the current cloud IP pool allocation strategy. In response to our disclosures and their own internal review of customer's configurations, Amazon is updating best practices to advise customers and providing additional guidance within the user experience of certain services (\autoref{deployed_mitigations}).

\section{Background}\label{background}
We study the prevalence of security risks associated with cloud-application configuration. It lies at the intersection of three bodies of work: network telescopes, architecture of public clouds, and configuration management. We provide background on each, then discuss the problems surrounding IP reuse in public clouds that motivate our measurement study.

\subsection{Internet Telescopes}
Internet telescopes are systems designed to observe network events for large-scale analyses~\cite{antonakakis_understanding_nodate,durumeric_search_2015,wustrow_internet_2010,bailey_internet_2005}. Telescopes vary in design and purpose but serve to capture a representative sample of events from a particular vantage point in the network. The telescope can either interact with sources per some identified protocol or be non-interactive (\ie, passively collect observable traffic). The captured data (e.g., IP packets) can then be used to inform network health monitoring and intrusion detection systems, among other network services.

For our study, we design a large-scale telescope that provides visibility into cloud networks. We leverage publicly available offerings to provision virtual servers and collect inbound traffic for analysis. Notably, the telescope is designed to be non-interactive above the transport layer (e.g., TCP).

\subsection{Architecture of Public Clouds}
Public cloud infrastructure drives a myriad of services offered by large corporations---ranging from productivity suites~\cite{protalinski_microsoft_2010} to managed blockchains~\cite{geuss_ibm_2016}. There are several models by which (public) cloud service providers lease out shared resources to tenants, including Software-as-a-Service (SaaS), Platform-as-a-Service (PaaS), and Infrastructure-as-a-Service (IaaS). Unlike SaaS and PaaS, where the tenant largely delegates management of hardware and infrastructure to the cloud service provider, IaaS customers directly manage their virtualized storage, network, and compute resources. Here, we focus our efforts on the most fundamental option offered under an IaaS model: virtual private servers. Amazon Web Services (AWS) offers virtual private servers primarily through their Elastic Compute Cloud (EC2) platform~\cite{wang2010impact}.

Abstracting compute, network, and storage resources provides the illusion of a private cloud. Cloud tenants can use APIs to create virtual servers, which can be assigned public IP addresses and accept traffic from other services within the cloud, or from the Internet at large. Regardless of which model a tenant leases resources under, individual service endpoints are typically uniquely identified via distinct IPs. The IPs are provisioned to the tenant from the cloud provider's pool of IPs, and can be reused by others once the tenant releases them. As different tenants continuously provision and release IPs from the shared pool, the risk of inconsistencies in configuration settings increases.

\subsection{Configuration Management and Latent Configuration}
Traditionally, enterprise compute, storage, and network resources were kept on-premises and largely managed with a collection of scripts and ad hoc practices. However, outsourcing to a cloud-service provider introduces a radically new interface through which tenants deploy and manage network services. The usage model typically follows an \textit{allocate-run-scale-deallocate} pattern that defines the characteristic \textit{elastic} property of public clouds~\cite{herbst2013elasticity}. While favorable for the scalability and reliability benefits, managing large-scale systems on elastic cloud resources is an increasingly difficult task~\cite{dillon_cloud_2010}. Specifically, tenants are responsible for correctly decommissioning all related services, configuration settings, and other dependencies during the \textit{decommissioning} phase of operation.

Fundamental to our work is the concept of configuration settings not removed during decommissioning, which we refer to as \textit{latent configuration}. Concretely, we focus our study on IPs in the cloud that a tenant has released and no longer controls, but failed to remove the (now latent) configuration referring to it. It is latent because, while it initially does not refer to a valid resource and is therefore harmless, another (potentially adversarial) tenant could receive the IP address it refers to, making the configuration active and potentially exploitable. Work studying the issue of latent configuration~\cite{xu_early_nodate} has generally shown it to be a difficult risk to mitigate, as it initially causes silent failures, evading detection by system administrators and security organizations. 

\subsection{IP Reuse and Dangling DNS}\hypertarget{mr2a}{}\hypertarget{mr7a}{}
IP Reuse is an inherent property of public cloud architectures. Liu et al.~\cite{liu_all_2016} performed an investigation of this issue in the context of the DNS, wherein \textit{dangling DNS records} pointing to an IP address or other resource could allow an adversary to take over domains by controlling the reused resource. The authors additionally found dangling DNS records exploitable beyond IP reuse when records point to other hosted platforms (\eg, GitHub pages or Heroku). Others have demonstrated additional generality, as nameserver (\texttt{NS}) records can allow domain takeovers~\cite{alowaisheq_zombie_2020}. An adversary performs a squatting attack by acquiring the referenced resource, whether that be an IP address or other service identifier. Indeed, further large-scale studies demonstrated that dangling DNS vulnerabilities are prevalent and span beyond IP reuse~\cite{noauthor_dangling_2021}.

In a related work, Borgole et al~\cite{borgolte_cloud_2018} further investigated dangling DNS and IP reuse specifically. They found that APIs for bulk provisioning of IPs often allow an adversary to acquire a given IP address associated with a target domain. They further studied the effect of existing and proposed security techniques such as TLS towards preventing exploitation of dangling DNS records. While a properly issued and used TLS certificate guards against the effects of dangling DNS records, the use of domain validation during certificate issuance (e.g., through LetsEncrypt~\cite{noauthor_lets_nodate}) allows an adversary to provision a valid certificate using the dangling DNS record, defeating this protection. They proceed by demonstrating defenses against the domain-validation vulnerability through an enhanced domain ownership challenge.

These works all find vulnerable web properties through scanning of DNS, and security teams within organizations have responded by using domain scanning tools to detect and mitigate such vulnerabilities. However, DNS is but one means by which latent configuration can occur. We investigate vulnerabilities caused by IP reuse directly, finding classes of vulnerabilities that are independent of DNS and that cannot be detected by subdomain scanning techniques.

\section{Cloud Squatting}\label{about}

\begin{figure}[h]
\centering
\includegraphics[width=0.7\columnwidth]{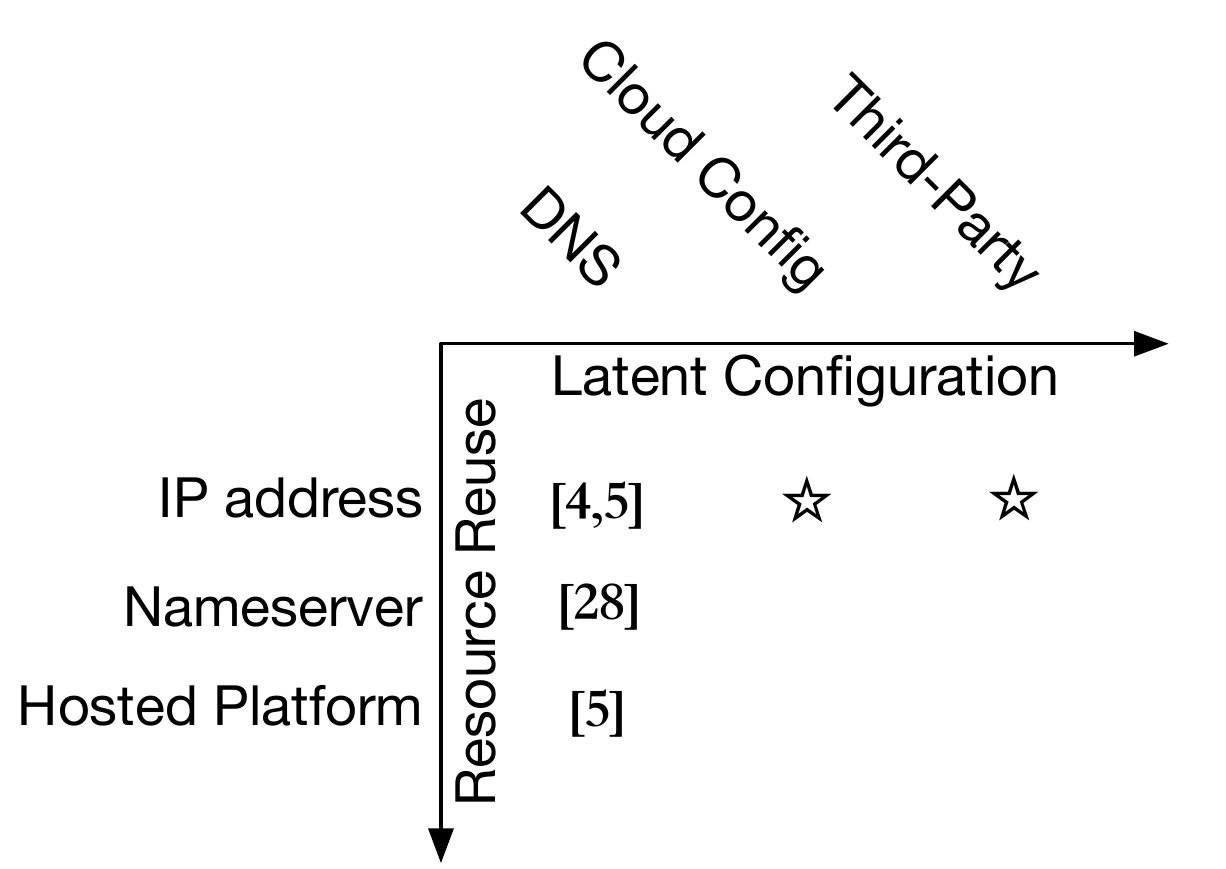}
 \caption{The space of cloud squatting threats. New attacks explored by our work are represented by a star.}
\label{fig:threat_space}
\end{figure}

In this work, we study latent configuration and resource reuse in clouds generally. To date, this vulnerability has been considered in the context of dangling DNS and IP use-after-free. Common amongst these vulnerabilities and our considered space is that vulnerabilities are exploited via squatting on the resource, a concept that has received ongoing attention in the security community (\eg, combo squatting~\cite{kintis_hiding_2017}, typo squatting~\cite{szurdi_long_2014}, file squatting~\cite{vijayakumar_sting_2012}, and skill squatting~\cite{kumar_skill_2018}, among others). Motivated by the generality between these topics, we name this superset attack space \textit{cloud squatting}. An adversary can perform a cloud squatting attack when latent configuration refers to a cloud resource (e.g., an IP address) that the adversary controls (see \autoref{fig:overview}). Though we focus our study on configurations referring to IPs, it is conceivable that myriad cloud configurations referring to resources other than IPs could also exist and be exploitable (\autoref{fig:threat_space}). We next discuss the general classes of latent configurations an adversary might seek to exploit, as well as the steps they would take to exploit it.

\subsection{Classes of Latent Configuration} 

While latent configurations are as diverse as the applications deployed on public clouds, we characterize them into three classes: cloud services, third-party applications, and DNS. Each represents unique properties and challenges for study.

\sblock{Cloud Services.} Cloud providers offer a selection of services that can be used in tandem to ease development, deployment, and management. Tenants typically opt to use these services because of the level of automation they can achieve---for fault tolerance, scalability, and more~\cite{noauthor_building_2017}. While each cloud provider offers different options, commonly used services include content distribution networks, pub/sub messaging systems, network monitoring, and API servers. AWS's offerings include the CloudFront~\cite{cloudfront} content distribution network, the Simple Notification Service~\cite{sns} (SNS) message queuing service (similar to MQTT~\cite{mqtt}), Route53 health checks~\cite{route53} that monitor server/service status, and API Gateway~\cite{api_gateway} that provides a frontend for API requests to instances in the cloud.

Services are largely provisioned with an accompanying IP address, which is used for communication. IP addresses are assigned to new virtual servers from a shared IP pool, and cloud tenants then set up service associations between them (\ie, which service endpoints communicate with each other). Under public cloud providers such as AWS, when tenants launch virtual servers for public web services, the servers can receive both public IPs as well as public DNS hostnames formed using deterministic IP address based naming (IPBN~\cite{aws_dns_hostnames}, e.g., \texttt{ec2-203-0-113-15.compute-1.amazonaws.com}). As a result, references to these raw resource identifiers can be stored as DNS records or directly in service configurations. Failing to remove configurations pointing to raw resource identifiers when decommissioning the servers then leads to latent configuration.

For example, for SNS, tenants specify which endpoints (per their fixed IP based hostname) should receive notification messages for different SNS topics. The configurations between the services, however, are not automatically removed when the tenant releases the public IP address. If the configuration persists after the tenant releases the IP, SNS will continue to send messages to the same IPs---even if the IP is subsequently assigned to an instance now owned by an adversary. 

\sblock{Third-party applications.} Just as cloud services use raw resource identifiers, applications deployed by tenants often can as well. For instance, a webserver may be configured to connect to a database to store transaction information. When this configuration references the database via an IP address or IPBN, there is the potential for latent configuration to be created. If an adversary subsequently creates a virtual server and receives the IP address associated with the database, the previous tenant's webserver will continue to connect to the IP and thus risk leaking sensitive data to the adversary.

This issue is exacerbated by the use of services that are fault tolerant: an application that silently tolerates failures due to latent configuration may also silently resume connecting to the service when it is controlled by an adversary. Because these services are developed or deployed by each tenant individually, it is difficult to fully quantify the space of these vulnerabilities. We therefore discuss classes of vulnerable third-party services deployed by tenants in \autoref{exploiting:other}.

\sblock{DNS.} The ubiquity and transparency of DNS makes it an especially compelling target for study, as demonstrated by the prior works that have measured incidence of dangling DNS. Clients use DNS to resolve human-readable names to IP addresses. Most importantly, end users place trust in the associations provided through DNS, giving sensitive information to organizations based on DNS records. When cloud tenants decommission servers and leave DNS configuration referencing these servers, user- or machine-driven clients can connect to these servers and trust them to be associated with the original organization. As analyzed in \autoref{exploiting:dns}, organizations of all sizes leave DNS records referring to servers they no longer control. Configuration errors in DNS have received substantial consideration from the security community, with takeovers of domains being demonstrated at various levels of the resolution process~\cite{liu_all_2016, alowaisheq_zombie_2020, noauthor_amazon_2021, borgolte_cloud_2018}. However, being a publicly-accessible directory of configuration, DNS is a promising target for our study, as discovered vulnerabilities can be definitively traced back to specific organizations. DNS can also be used as part of the other studied configuration classes--for instance, a cloud service could resolve an IP through DNS. Our results (\autoref{exploiting}) show that, while this does occur in practice, much of the measured exploitable traffic was configured independently of DNS.

\subsection{Exploiting Latent Configuration}

Upon detecting latent configuration, the adversary must take steps to actually exploit any received traffic. In the case of DNS, as well as some cloud service traffic (\eg, Route53, Cloudfront), phishing attacks can be performed~\cite{liu_all_2016}, including creation of valid TLS certificates in some cases~\cite{borgolte_cloud_2018}. Other cloud service traffic (\eg, SNS and API Gateway) can directly send sensitive information to an adversary. Thus, an adversary need only retain the IP address or other resource and continue to (passively) receive sensitive data.

Under an untargeted attack, an adversary would first determine the owner of the sensitive data, then either solicit a ransom payment from that party or offer the data for sale online. Indeed, our experiment found instances of private information leakage similar to that involved in high-profile user data leaks~\cite{paz_t-mobile_2021}. Other areas of study have similarly established untargeted attacks as compelling for adversaries, such as phishing~\cite{dhamija_why_2006, hong_state_2012}, ransomware~\cite{kharraz_cutting_2015, laszka_economics_2017}, and data exfiltration~\cite{al-bataineh_analysis_2012,ullah_data_2018}. For service traffic, such as databases, the adversary could serve incorrect data to the clients, or store sensitive data that the client attempts to insert in the database. In some cases a sophisticated adversary could infer the service previously hosted on an IP, connect to the new IP address of the service, and perform a man-in-the-middle attack. 
\section{Cloud Traffic Collection}
We aim to demonstrate the existence, diversity, and prevalence of latent configuration vulnerabilities in a public cloud. Here, we describe our collection approach, followed by ethical and adversarial implications.

\subsection{Measurement Approach}

We aim to collect a representative sample of traffic inbound to public cloud IP addresses, most importantly, by maximizing the number of IP addresses observed. The approach is inspired by extant works on Internet telescopes~\cite{antonakakis_understanding_2017, wustrow_internet_2010, pang_characteristics_2004}, but differs in that it is deployed in IP space allocated to cloud providers, rather than dark space (\ie, portions of the IP address space not allocated for an active purpose). Collection servers are automatically allocated (using IP addresses from the AWS IP pool) until a quota is reached of simultaneous servers (in our case, the quota set by AWS was \SI{320}{} simultaneous servers). Each collection server accepts TCP connections on all ports and records received traffic for up to \SI{10}{} minutes. After \SI{10}{} minutes, the collection server uploads received traffic in PCAP format to an encrypted log repository (\ie, Amazon S3) and terminates. Terminated servers are automatically replaced by new collection servers to maintain the server quota. In this way, new IPs are continually drawn from the pool and information on connecting services is collected.

\subsection{Limitations}

Our approach to measurement of latent configuration takes a different approach from prior works, which analyze public configuration repositories such as DNS~\cite{borgolte_cloud_2018, liu_all_2016}. While this allows us to see vulnerable configurations that are not publicly visible, it also carries new considerations and limitations: (1) In some protocols (\eg, Postgres), the server is expected to execute a specific protocol for the client to continue sending data, so our passive telescope can only see limited data on these protocols. (2) Our approach only sees vulnerabilities that actively cause traffic to be sent during the 10-minute study period (this also means observed latent configurations are more likely to be in active use). In these ways, our measurement approach complements those of prior works.

\subsection{Ethics and Adversarial Implications}
\hypertarget{mr1a}{}
Throughout our study we took actions to ensure that effects of our measurement would be minimized. As in prior works~\cite{borgolte_cloud_2018, liu_all_2016}, we capped our IP addresses allocation to an acceptable rate (320 addresses every 10 minutes). Our data collection was covered under an exemption from our institutional review board (IRB). While the scope of data collected in our study was similar to other network telescopes~\cite{wustrow_internet_2010, antonakakis_understanding_2017,bailey_internet_2005}, we also took additional steps (outlined in \refappendix{appendix:data_handling}) to ensure that data was protected throughout the study. Disclosure of all discovered vulnerability was performed through Amazon (\refappendix{appendix:disclosure}), including extended scanning by AWS to provide expanded disclosure.

Unlike previous datasets that collect only transport-layer traffic (\eg, UDP packets and TCP \texttt{SYN} packets)~\cite{richter_scanning_2019}, our approach yields raw packet captures with data from servers that are legitimately routable, but otherwise have no content. Because the approach does not rely on privileged access to the cloud, it also presents a compelling technique for an adversary: rather than passively collect traffic for study, an adversary could deploy honeypots designed to target commonly-used protocols. These honeypots could record personal information for exploitation, provide fake authentication prompts to extract credentials, or host drive-by downloads of malware. The low cost with which our measurement study was performed (\SI{2089,76}{USD} over \SI{101}{} days) suggests that an adversary could carry out such an attack at minimal expense. This clear risk to cloud infrastructure motivated our extensive disclosure and remediation process (see \refappendix{appendix:disclosure}).
\section{Characterizing Cloud IP Use}\label{cloudip}

\begin{table}[]
    \caption{IP allocation statistics, including per-region estimates of the total available IPs and percentage of estimated IPs that were measured in our study (Capture Rate). In total we estimate that $\SI{56}{\percent}$ of available IP addresses in the \texttt{us-east-1} region were measured.}
    \centering
    \begin{tabular}{ccccc}
        \toprule
        Zone & Servers & Unique IPs & Estimated IPs & Capture Rate \\
        \midrule
us-east-1a & $\SI{581}{k}$ & $\SI{383}{k}$ &  $\SI{789}{k}$ &  $\SI{49}{\%}$ \\
us-east-1b & $\SI{607}{k}$ & $\SI{389}{k}$ &  $\SI{762}{k}$ & $\SI{51}{\%}$ \\
us-east-1c & $\SI{630}{k}$ & $\SI{236}{k}$ &  $\SI{313}{k}$ & $\SI{76}{\%}$ \\
us-east-1d & $\SI{573}{k}$ & $\SI{360}{k}$ &  $\SI{700}{k}$ & $\SI{51}{\%}$ \\
us-east-1f & $\SI{647}{k}$ & $\SI{171}{k}$ &  $\SI{198}{k}$ & $\SI{87}{\%}$ \\
\hline
Total & $\SI{3039}{k}$ & $\SI{1540}{k}$ &  $\SI{2762}{k}$ &  $\SI{56}{\%}$ \\
        \bottomrule
    \end{tabular}
    \label{tab:estimated_ips}
\end{table}

We first use our collected data to analyze the AWS \texttt{us-east-1} IP pool. An adversary wishing to exploit latent configuration would aim to measure as many IPs as possible, and to ensure that those IPs have been used by other tenants recently. This motivates two analysis questions: \textbf{(A)} how many IPs are available for allocation by cloud tenants? \textbf{(B)} how quickly are IPs available for reuse? These questions will also inform our evaluation of countermeasures (\autoref{sec:mitigations}).

\subsection{IP Address Availability}

To estimate the number of available IPs, we model the IP address pool in each AWS availability zone as a population survey. Population surveys are a statistical method generally used to measure animal populations, but the same principles can be applied in this case to estimate size and activity of the IP address pool. We begin by assuming that IP allocations are pseudo-randomly drawn from the pool of available IPs (as has been subsequently confirmed in conversations with Amazon). We model the pool as an open population, since other tenants also allocate and return IPs during the course of the study. All modeling was performed using an open population estimation technique developed by Sandland and Cormack~\cite{sandland_statistical_1984}, implemented in \texttt{Rcapture}~\cite{baillargeon_rcapture_2007}. We see that larger availability zones yield largely unseen IPs throughout the experiment whereas smaller ones are quickly covered. An adversary seeking to maximize IP coverage might target zones with fewer IPs, while one searching for a specific tenant's IPs would emphasize high capture rate.

Results of our population estimation are shown in \autoref{tab:estimated_ips}. We estimate the number of IPs in the pool at any point during the study, as well as the \textit{capture rate}, which is the percentage of estimated IPs that we measured. This can be interpreted as a probability that any IP released into the pool was measured by us during the study. We conclude that the current IP pool implementation on AWS is favorable for achieving high coverage of the IP space. Creating servers on AWS yields a high number of IPs, each of which could have potential latent configuration. Further, our capture ratio across each zone was as high as \SI{87}{\percent}, meaning that an IP released by a tenant in the pool had an \SI{87}{\percent} chance of being measured by our study. These metrics show that an adversary can continually measure the IP space and discover new, potentially exploitable systems, and that even a single adversary performing such an attack poses a high risk to cloud tenants in even the largest zones.

\subsection{Age of IPs at Reuse}

\begin{figure}
    \centering
    \includegraphics[width=\columnwidth]{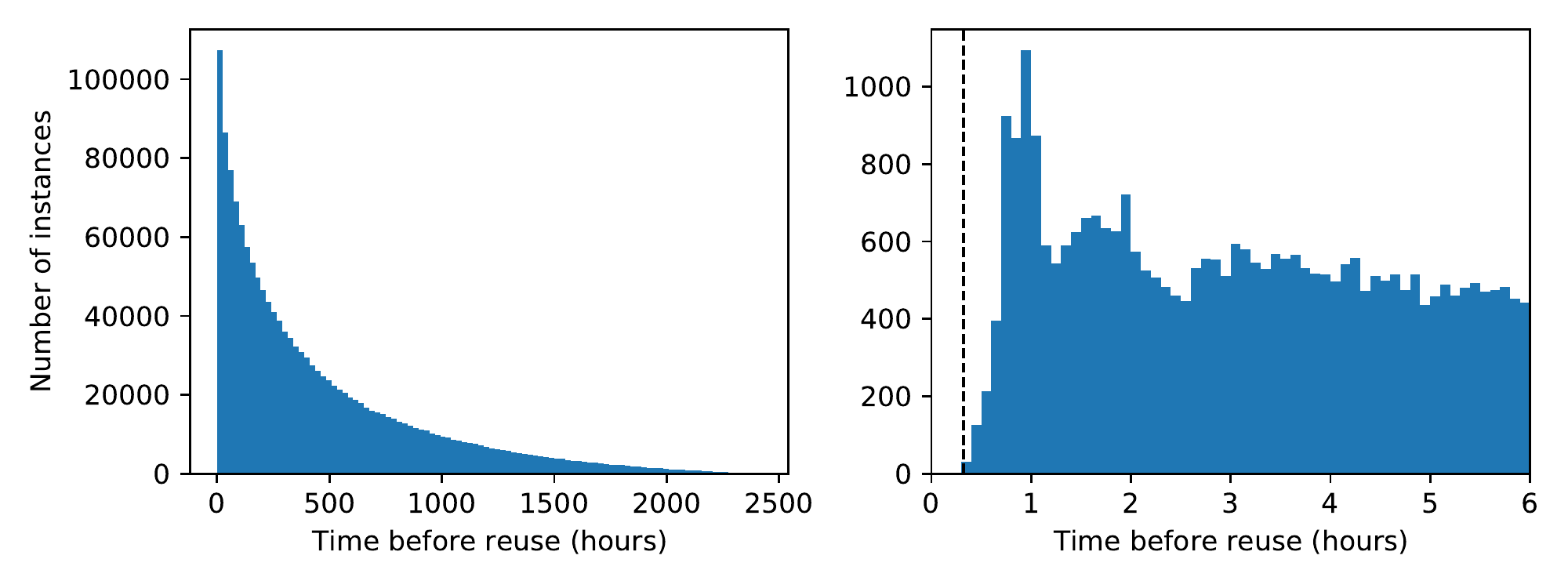}
    \caption{Measuring time between IP reuse on AWS, over the entire study and reuse seen within 6 hours. IPs were generally not reused within 30 minutes after release.}
    \label{fig:reuse_time}
\end{figure}

We additionally evaluate the age of IP addresses when they are reused (\ie, how long it takes for an IP address to be reallocated after a tenant releases it). Because we achieve such high coverage of the \texttt{us-east-1} IP space, many of the IPs seen by our apparatus are seen twice or more (in one instance in \texttt{us-east-1f}, we received the same IP address \SI{13}{} times). By recording the interval between release and reacquisition of these IP addresses, we can characterize the IP address allocation to understand what policies are being applied.

\autoref{fig:reuse_time} shows the distribution of these intervals. At a macro level (left side), we see that reuse time is consistent with a Poisson process, implying that IP addresses are randomly chosen from the pool without respect to when the IP was most recently used. When looking at reuse seen within \SI{6}{} hours (right side), we find that IP addresses are not reused within \SI{30}{} minutes of release, even by the same tenant (outliers are caused by our use of EC2 spot instances and do not indicate reuse before \SI{30}{min}). This demonstrates AWS is employing a cooldown policy on their IP address pool, though as we see in our analysis of exploitable configuration (\autoref{exploiting}) such an aging policy is likely intended for other purposes and does not prevent exploitable misconfiguration due to IP reuse.

\section{Measuring Exploitability}\label{exploiting}

Referring back to \autoref{tbl:vulnerability_taxonomy}, we consider latent configuration vulnerabilities in three stages. We first explore vulnerabilities associated with cloud services. Next, we show that vulnerabilities exist beyond managed services and span a variety of protocols, applications, and verticals. Finally, we leverage information obtained from public DNS to attribute found vulnerabilities to specific organizations, finding that latent configuration is ubiquitous across organizations of all sizes.

\subsection{Exploitation Through Cloud Services}\label{exploiting:cloud}

We first investigate vulnerabilities caused by latent configuration in managed cloud services. To do this, we filtered all received traffic to sessions exclusively coming from the AWS IP space that is reserved for managed services~\cite{aws_ip_ranges}. Requests were then associated with each individual cloud service based on HTTP user agent. Note that, because the studied IP space is reserved for managed services, we can validate that filtered traffic is legitimate service traffic and not an adversary posing as a managed service.

In total we discovered traffic that was traceable to four different cloud services: (1) Simple Notification Service (SNS), (2) Route53, (3) Cloudfront, and (4) API Gateway. While these services varied in prevalence, each service either directly sent sensitive data or had a clear path by which an adversary could extract sensitive data. Coarse statistics on traffic received from each service are presented in \autoref{tab:cloud_services}.

\begin{table}[]
    \caption{Number of unique IPs receiving traffic from each cloud service, number of TCP sessions in total and with DNS info, and estimated unique tenants.}
    \centering
    \begin{tabular}{rcccc}
        \toprule
        \multicolumn{1}{c}{Service} & \multicolumn{1}{c}{SNS} &  \multicolumn{1}{c}{Route53} &\multicolumn{1}{c}{Cloudfront} &  \multicolumn{1}{c}{API Gateway}\\
        \midrule
        IPs &\SI{24.9}{k} & \SI{2.8}{k} & \SI{65}{} &   \SI{3}{} \\
        Sessions & \SI{1.6}{M} & \SI{3.6}{M} & \SI{1.7}{k}  & \SI{10}{}\\
        Sessions w/ DNS & \SI{25}{} & \SI{567}{k} & \SI{767}{}  & \SI{2}{}\\
        Unique Tenants & \SI{78}{} & \SI{3.1}{k} & \SI{64}{} & \SI{3}{}\\
        \bottomrule
    \end{tabular}
    \label{tab:cloud_services}
\end{table}

\sblock{SNS.} AWS Simple Notification Service (SNS)~\cite{sns} allows tenants to publish and subscribe to messages (similar to MQTT~\cite{mqtt}) and is broadly used for internal communication between cloud services. It is designed to be fault tolerant: it will continue sending messages to subscribed IP addresses even if the address is not available for an extended period of time, enqueueing failed messages as necessary (\eg, due to server decommissioning). Therefore, an adversary who receives the IP address may receive new messages and messages enqueued from before they acquired the address.

We received messages from \SI{78}{} SNS topics on \SI{24.9}{k} unique cloud IP addresses, with \SI{1.6}{M} total messages received. Because SNS traffic is intended for internal communication between services, some of the communications received from these channels were highly sensitive. In one case, a SNS endpoint was used by a financial services provider to transmit information pertaining to client transactions. In another, a social services organization was transmitting the names, addresses, contact information, and location history of clients via SNS. In both of these cases, latent configuration referenced multiple previously-controlled IP addresses for a single topic, amplifying the ability for an adversary to receive the traffic.

\sblock{Route53.} Route53~\cite{route53} is Amazon's authoritative DNS service, and it allows tenants to check the health of services (Route53 Health Checks) before routing traffic to them. The health checks also provide fault tolerance: an unreachable service will not be included in DNS responses, preventing traffic from reaching that service. However, if a health check targets an IP address now-owned by an adversary, they can begin responding successfully to the health check and subsequently receive traffic intended for the previous tenant. 

Although this traffic did not directly contain sensitive data, it is indicative of a cloud squatting vulnerability. Most health checks were not associated with a domain name, however, in some cases health checks were associated with domain names that were also seen directly receiving end-user traffic. For instance, one entertainment company had vulnerabilities under multiple unique domain names and under Route53 health checks. In total \SI{2.8}{k} unique IP addresses received traffic from Route53 health checks, with an estimated \SI{3.1}{k} unique properties (this implies that some IP addresses were associated with multiple properties, potentially by the same tenant).

\sblock{Cloudfront.} Cloudfront~\cite{cloudfront} is a content delivery network (CDN) that routes requests to cloud servers. When a tenant fails to remove an IP address from Cloudfront configuration when releasing the IP address, a cloud squatting vulnerability can occur. In total we found \SI{65}{} IP addresses received \SI{1.7}{k} requests from Cloudfront. Because Cloudfront distributions are often placed in front of static content we predictably did not observe sensitive data sent directly by these connections. However, the use of CDNs in serving content such as scripts makes them an enticing target for adversaries. In one case, a Cloudfront distribution forwarded requests for a JavaScript file to be run alongside a major website. An adversary responding to this request could receive remote code execution capabilities within the context of a trusted site.

\sblock{API Gateway.} API Gateway~\cite{api_gateway} also acts as a frontend to route traffic to cloud services. We found that three unique IP addresses received traffic attributable to API Gateway. In one case, this traffic directly contained API authentication information intended for a service hosted on AWS. While the limited set of connections from this service makes it difficult to draw conclusions on trends, it is likely that cloud squatting presents a credible concern for customers using API Gateway with raw IP addresses.

\sblock{Independence from DNS.}\hypertarget{mr5a}{} We confirmed our hypothesis that latent configuration might exist beyond just DNS: across the 4 cloud services measured, traffic existed on each that did not contain DNS information. Because this traffic is sent from a managed cloud service with standardized behavior, the lack of DNS information in a request implies that DNS was not used to configure the connection from the cloud service (\ie, the configuration directly referred to the IP address). As a result, we conclude that the connections are caused by latent configuration within the cloud service itself. In these cases, as cloud provider configuration is not publicly visible, our telescope-based approach identifies vulnerabilities not seen in prior works.

\sblock{Generality across providers.} Our measurement study is focused on the specific services provided by AWS. As other cloud providers offer similar services, we hypothesize similar effects on those providers as well. For instance, Microsoft Azure offers Azure Event Grid~\cite{azure_event_grid}, which is similar to SNS, and Google Cloud Platform offers Cloud Pub/Sub~\cite{gcp_pubsub}. In the case of Azure, endpoints are validated at provisioning time, but configuration is not validated on an ongoing basis. Cloud Pub/Sub relies on TLS for endpoint authentication, so findings regarding TLS apply here as well.

\vspace{6pt}

Our analysis demonstrates a trend of exploitable latent configuration across managed cloud services. Whereas an adversary wishing to exploit connections through a CDN or health check must exert manual effort to perform a phishing campaign, messages received from a service such as SNS directly convey sensitive information, and an adversary could obtain this traffic in a fully automated fashion, analyzing the data after the fact to determine what information is of value. Further, these cloud services collectively show a surprising downside of fault tolerance: when services fail silently and recover from errors automatically, they can inadvertently become targets for abuse by adversaries when used in a shared compute environment such as a public cloud. While our study identified many vulnerable properties, the approach only allows us to observe a (representative) fraction of total cloud traffic. In response to observed vulnerabilities, Amazon is using control-plane information to detect potential vulnerabilities across all regions and tenants (\autoref{deployed_mitigations}).

\subsection{Exploitation through Third-Party Services}\label{exploiting:other}

\begin{table}[]
    \caption{Effect of the traffic filtering apparatus on remaining traffic to be analyzed. Our goal in this section is to reduce the dataset to be manually analyzed for exploitable traffic.}
    \centering
    \begin{tabular}{cccc}
        \toprule
        Step & IPs & TCP Sessions & Size\\
        \midrule
        Initial & \SI{3.13}{M} & \SI{596}{M} & \SI{410}{GB} \\
        Network & \SI{3.03}{M} & \SI{280}{M} & \SI{148}{GB} \\
        Transport & \SI{1.70}{M} & \SI{10.2}{M} & \SI{11}{GB}\\
        Session & \SI{1.14}{M} & \SI{4.89}{M} & \SI{9.3}{GB}\\
        Application & \SI{340}{k} & \SI{2.95}{M} & \SI{6.3}{GB}\\
        \bottomrule
    \end{tabular}
    \label{tab:filtering}
\end{table}

We filter collected traffic to examine the prevalence of latent configuration in third-party service traffic (\ie, traffic not sent by managed cloud services). Our filter (see sidebar and \refappendix{appendix:funnel}) attempts to identify what is legitimate (\ie, attempting to interact with a previous tenant of the IP address) and exploitable (\ie, allows for unintended information leakage from the client).

\begin{sidebar}

\section*{Filtering Process}

\begin{figure}[H]
    \centering
        \centering
        \includegraphics[width=\textwidth]{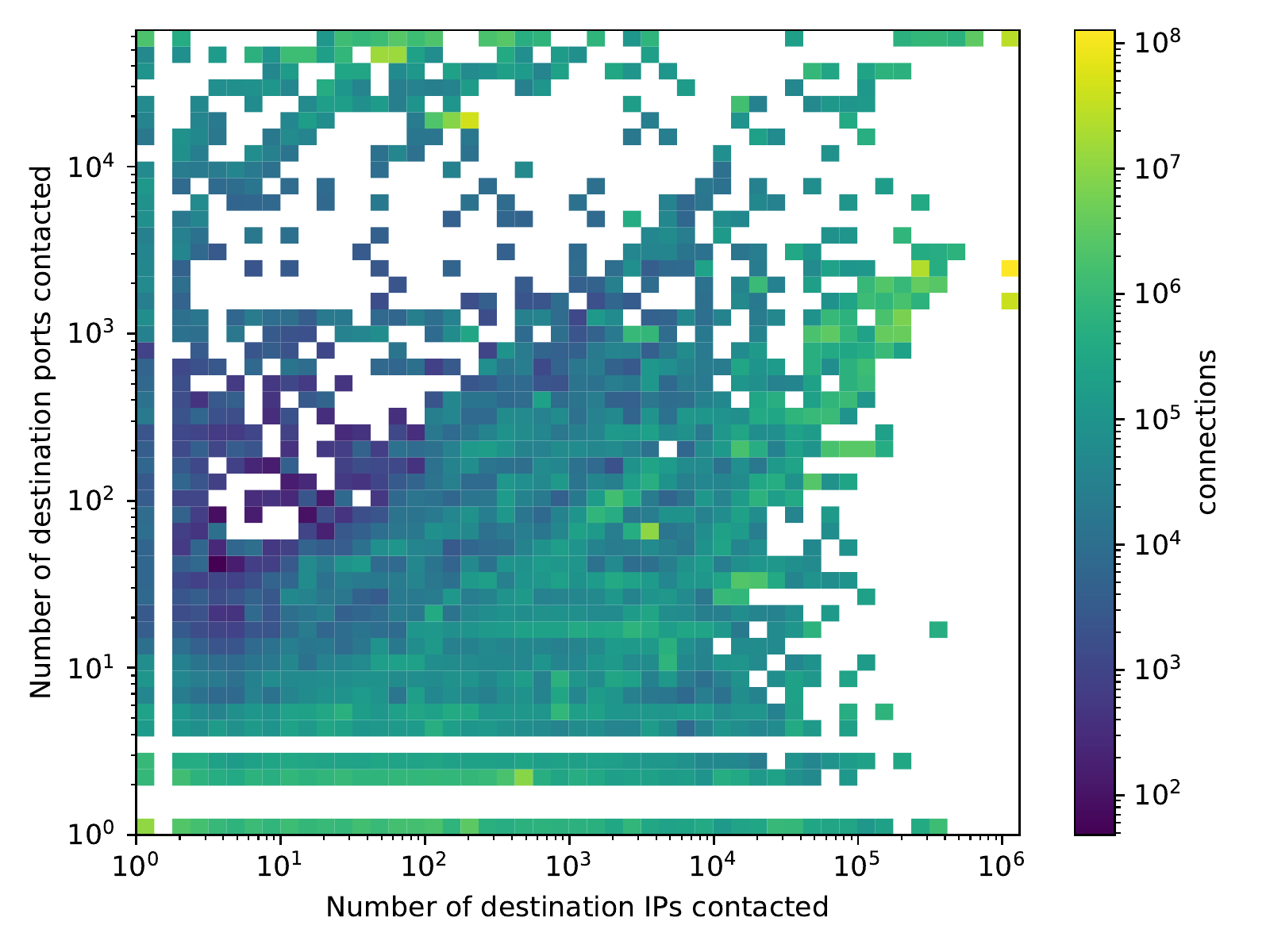}
    \caption{Ports and IPs contacted by each source IP. Only traffic under the bottom left data point is retained.}
    \label{fig:no_dups}
\end{figure}

We briefly describe the components of our filtering apparatus, which consists of four steps (results of each filtering step on dataset size are shown in \autoref{tab:filtering}):
\vspace{3pt}
\sblock{Network.} Coarse network filtering eliminates obviously non-legitimate traffic by filtering against several publicly-available blocklists~\cite{tsaousis_firehol_nodate}. Blocklists have the advantage that they have generally been independently evaluated, and are intended to have low false positives.

\sblock{Transport.} Consider the IPs and ports on our servers accessed by each source IP. We filter out sessions from source IPs that contacted either multiple of our IP addresses, or multiple ports, as this is likely scanner or other exploit traffic. As shown in \autoref{fig:no_dups}, the majority of traffic is sourced from IP addresses that exhibit scanning behavior. While this filters out legitimate traffic (\eg, cloud service traffic), it is an acceptable trade-off to reduce the size of our dataset.

\sblock{Session.} Clients that fail to complete the TCP handshake process or that do not send any data upon connecting are likely scanner traffic and are therefore also ignored. Much of this traffic is associated with distributed botnets~\cite{antonakakis_understanding_2017, bailey_survey_2009}. In other cases, the traffic may be legitimate and exploitable, but uses a protocol (such as Telnet) that does not initially send a payload. Such protocols are not as readily studied by our approach and can therefore be excluded.

\sblock{Application.} We use network intrusion detection rules from Snort~\cite{roesch_snort_1999} to further eliminate bot traffic, as well as additional manually-generated rules to filter peer-to-peer (\eg, BitTorrent and Bitcoin traffic that does not trust connected peers) and other cloud traffic that is likely not exploitable.

\end{sidebar}

After all filtering steps were applied, the remaining traffic contained \SI{340}{k} source IP addresses across \SI{2.95}{M} TCP sessions. The first session from each source IP address was analyzed manually to further reduce dataset size. While quantifying the prevalence of other exploitable traffic is intractable, we did find that exploitable traffic exists across a variety of protocols and applications.

\sblock{Databases.} We received traffic intended for databases hosted by customers on AWS. Connections were specifically identified across MySQL~\cite{mysql}, Postgres~\cite{postgres}, and ElasticSearch~\cite{elasticsearch}, though other database protocols likely exist within the data but were not manually identified. For example, one IP address received repeated connection attempts to a Postgres database apparently intended to hold payment information, including plausible credentials. An adversary could employ a database honeypot to harvest credentials for use in attacking other services, or even directly receive sensitive user data contained in database queries, though our experimental apparatus purposefully does not elicit such traffic.

\sblock{Caches.} We identified two types of traffic intended for Redis~\cite{redis}: (1) client traffic sending queries to an instance and (2) communication between Redis cache servers. While client traffic may be attributable to scanners, inter-server traffic was plausibly legitimate. This traffic implies that the IP address was formerly part of a cluster of Redis cache servers serving the same cache, and that a server listening at this address could receive intra-cluster communication traffic, which would contain data about the information stored in the cache. Again, the passive nature of our data collection approach did not allow sensitive information to be received in this case.

\sblock{Financial Traffic.} We identified an instance of traffic sent using the Financial Information eXchange (FIX) protocol~\cite{fix_protocol}. This protocol is used for sending metadata and commands related to securities trading. Manual analysis definitively traced this FIX traffic to a financial services startup: the organization likely previously hosted a FIX service at the IP address, but failed to remove all latent configuration when the service was decommissioned. The traffic contained credentials and further interaction would likely result in receiving information on transactions. We separately sent disclosures about this incident in addition to those discussed in \refappendix{appendix:disclosure}.

\sblock{Logging and Metrics.} We found many instances of mobile, web, and other applications sending logging, crash report, and metrics traffic to controlled IP addresses. These log requests contained tracebacks of application errors from mobile devices, metadata on device characteristics, and personal information. In some cases log entries were traceable back to specific websites or applications. For example, an online recipe site inadvertently sent analytics messages, including email IDs, viewed articles, and user IP address. Another IP, apparently associated with an advertising metrics service, received end-user device IMEI info (against best practices in smartphone privacy \cite{wetherall_privacy_2011}). An adversary allocating IP addresses in the cloud would receive such traffic automatically, and could collect personal information for use in spearphishing campaigns.

\sblock{Webhooks.} In the same way that cloud services can send messages directly to IP addresses, third-party hosted services could be configured to communicate with IP addresses that are no longer controlled. We observed traffic from Teamwork~\cite{teamwork}, BitBucket~\cite{bitbucket}, and Segment~\cite{segment} targeted at cloud IP addresses, likely caused by tenants who previously configured these IP addresses to receive notifications when events occurred in these platforms. In the case of Segment, for instance, webhook traffic contained recipient email addresses. Traffic from a Teamwork webhook apparently contained user communications from the platform. Webhooks are a convenient way to integrate custom software with third-party services. Yet, our study suggests that the information within these webhook messages may be intercepted by adversaries due to latent configuration. While tenants must ensure their webhooks are removed when receiving services are decommissioned, service providers also have an opportunity to protect customers by eliminating sensitive information from messages.

\sblock{Tenant APIs.} We additionally found traffic from applications attempting to communicate with APIs hosted by previous tenants. These endpoints could have been hardcoded into application source code or configured through some other channel. In one case, we observed what appeared to be search query autocomplete traffic from a notable third-party mobile browser (8509 such requests to a single IP address). Attempts were not made to decode this traffic for ethical reasons.

\sblock{Independence from DNS.} While the varied nature of third-party services makes this class of vulnerabilities difficult to quantify, third-party service traffic that does contain DNS information can be used to establish bounds on DNS independence. Of the \SI{2.95}{M} TCP sessions observed after filtering, \SI{1.20}{M} contained HTTP or TLS requests. Of these, \SI{970}{k} contained host headers. This header either directly contained an IP address (\SI{398}{k}/\SI{41}{\percent}), contained an IPBN (\SI{283}{k}/\SI{29}{\percent}), or contained other data such as a domain name (\SI{288}{k}/\SI{30}{\percent}). Presence of IPs and IPBNs in these host headers suggests that these connections (as much as \SI{70}{\percent}) were not configured through some other DNS-based configuration. Note that non-HTTP/TLS traffic may exhibit different behavior, and many other protocols (\eg, Postgres) do not send host information on connections, preventing analysis.

\vspace{6pt}

The variety of protocols and applications observed suggests that latent configuration through third-party services is a widespread problem. Such vulnerabilities can provide an adversary with traffic intended for internal use within applications, and can therefore disclose data that is highly sensitive. Further, the use of honeypots targeting common third-party services could likely extract this information in a fully-automated fashion across a high percentage of cloud IPs at minimal cost. Our analysis also suggests that much of this traffic is configured independently from DNS.

\subsection{Attributing Vulnerabilities through DNS}\label{exploiting:dns}
\hypertarget{or1a}{}

\begin{table*}
\caption{Observed exploitable domains in the top 1,000 by site ranking and by number of unique exploitable server instances.}
\centering
\begin{tabular}{|r|c|c|c||r|c|c|c|}
\hline
\multicolumn{4}{|c||}{\bf Top Domains by Unique Hosts} & \multicolumn{4}{c|}{\bf Top Domains by Site Ranking} \\
\hline
{\bf Site rank} & {\bf Unique hosts} & {\bf Subdomain Depth} & {\bf Domain} & {\bf Site rank} & {\bf Unique hosts} & {\bf Subdomain Depth} & {\bf Domain} \\
\hline
\hline
% The \makebox here is done to ensure both sides of the table are the same width
\SI{450}{} & \SI{107}{} & \SI{2}{} & redhat.com & \SI{31}{} & \SI{2}{} & \SI{4}{} & \makebox[\widthof{filemaker-cloud.com}][c]{amazonaws.com}\\\hline
\SI{8543}{} & \SI{72}{} & \SI{2}{} & splunk.com & \SI{68}{} & \SI{6}{} & \SI{2}{} & akadns.net\\\hline
\SI{1593}{} & \SI{65}{} & \SI{2}{} & service-now.com & \SI{76}{} & \SI{2}{} & \SI{2}{} & cnn.com\\\hline
\SI{588482}{} & \SI{57}{} & \SI{2}{} & filemaker-cloud.com & \SI{129}{} & \SI{1}{} & \SI{2}{} & wix.com\\\hline
\SI{76965}{} & \SI{47}{} & \SI{2}{} & boxcast.com & \SI{146}{} & \SI{2}{} & \SI{2}{} & harvard.edu\\\hline
\SI{11868}{} & \SI{34}{} & \SI{3}{} & appdomain.cloud & \SI{164}{} & \SI{33}{} & \SI{2}{} & go.com\\\hline
\SI{164}{} & \SI{33}{} & \SI{2}{} & go.com & \SI{177}{} & \SI{1}{} & \SI{1}{} & usatoday.com\\\hline
\SI{21057}{} & \SI{32}{} & \SI{1}{} & duckdns.org & \SI{284}{} & \SI{1}{} & \SI{1}{} & intuit.com\\\hline
\SI{7691}{} & \SI{25}{} & \SI{1}{} & hostedrmm.com & \SI{298}{} & \SI{1}{} & \SI{2}{} & cornell.edu\\\hline
\SI{7671}{} & \SI{22}{} & \SI{2}{} & yummly.com & \SI{300}{} & \SI{2}{} & \SI{1}{} & intel.com\\\hline
\SI{14349}{} & \SI{22}{} & \SI{1}{} & glance.net & \SI{302}{} & \SI{2}{} & \SI{1}{} & slack.com\\\hline
\SI{225042}{} & \SI{17}{} & \SI{1}{} & bitnamiapp.com & \SI{434}{} & \SI{1}{} & \SI{1}{} & vice.com\\\hline
\SI{291213}{} & \SI{16}{} & \SI{1}{} & qmetry.com & \SI{450}{} & \SI{107}{} & \SI{2}{} & redhat.com\\\hline
\SI{54293}{} & \SI{15}{} & \SI{1}{} & wostreaming.net & \SI{470}{} & \SI{4}{} & \SI{1}{} & trafficmanager.net\\\hline
\SI{2018}{} & \SI{14}{} & \SI{1}{} & ring.com & \SI{495}{} & \SI{1}{} & \SI{2}{} & upenn.edu\\\hline
\SI{65484}{} & \SI{14}{} & \SI{1}{} & otgs.work & \SI{497}{} & \SI{1}{} & \SI{2}{} & elsevier.com\\\hline
\SI{161178}{} & \SI{12}{} & \SI{2}{} & everlaw.com & \SI{535}{} & \SI{1}{} & \SI{1}{} & ieee.org\\\hline
\SI{226628}{} & \SI{12}{} & \SI{2}{} & reltio.com & \SI{578}{} & \SI{1}{} & \SI{3}{} & jhu.edu\\\hline
\SI{11565}{} & \SI{11}{} & \SI{3}{} & acquia.com & \SI{588}{} & \SI{1}{} & \SI{1}{} & nvidia.com\\\hline
\SI{16428}{} & \SI{10}{} & \SI{2}{} & psdops.com & \SI{618}{} & \SI{1}{} & \SI{3}{} & lenovo.com\\\hline
\SI{692115}{} & \SI{10}{} & \SI{2}{} & adikteev.io & \SI{767}{} & \SI{3}{} & \SI{3}{} & ea.com\\\hline
\SI{13518}{} & \SI{9}{} & \SI{1}{} & gannettdigital.com & \SI{782}{} & \SI{2}{} & \SI{1}{} & hhs.gov\\\hline
\SI{80657}{} & \SI{9}{} & \SI{1}{} & neulion.com & \SI{957}{} & \SI{1}{} & \SI{1}{} & justice.gov\\\hline
\end{tabular}
\label{tbl:topdomains}
\end{table*}

Recall, we allocate IP addresses and determine exploitable domains associated with these addresses based on requests from clients. In examining the contents of incoming requests, we can identify whom the requests are intended for (and thus intuit the likely application). Specifically, in a large number of sessions, banner information (the initial payload sent by the client upon connecting) includes the expected DNS hostname and identifies the expected service (owner). An adversary would use this information to pose as the service by creating a plausible looking interface for that domain or replicate an existing host (\eg, phishing websites).

We observed \SI{168}{} million client requests with DNS information in banner data. Of those, \SI{14309}{} were validated to real domain names (\ie, the request was received by the IP address that DNS actually resolved to), with \SI{719}{} unique domains listed in the top million (using Tranco~\cite{le_pochat_tranco_2019}). \hypertarget{mr6a}{}We additionally filtered domain names that were synonymous with IP addresses (\eg, IPBNs such as \texttt{ec2-203-0-113-15.compute-1.amazonaws.com} and wildcard DNS providers such as \texttt{xip.io}). We attribute the remaining sessions largely to botnet traffic and other background noise. We focus on the \SI{14309}{} sessions that mapped to valid DNS records, as they are of most likely value.

The sessions identified \SI{5446}{} unique second-level domains (SLDs, e.g., \texttt{example.com}), 495 of which had more than one unique subdomain identified.  The domains occurred in \SI{231}{} unique eTLDs (from the Mozilla Public Suffix List~\cite{public_suffix_list}) with \texttt{.com} representing \SI{63}{\percent} of the unique hostnames, \texttt{.net} representing \SI{6.2}{\percent}, \texttt{.org} at \SI{3.9}{\percent}, \texttt{.io} at \SI{3.2}{\percent}, and \texttt{.com.br} at \SI{2.7}{\percent} each. The remainder of the domains belonged to a mix of lesser-used ICANN-issued names (\eg, \texttt{.info}) country-coded TLDs (\eg, \texttt{.co.uk} and \texttt{.ru}), and gTLDs (\eg, \texttt{.cloud}). Note that because all data was collected from the \texttt{us-east-1} region, it may be biased towards domains related to organizations in the eastern side of the United States. Other regions will likely involve other domains and TLDs based on their geographic location and consumer populations.

Next, we investigated what kinds of domains may be vulnerable based on the banner data. \autoref{tbl:topdomains} shows domains based on two rankings. The left-hand side of the table ranks the domains that had the most unique hostnames observed in incoming requests. Here several of the highest ranking domains were from organizations that provide custom workflows or information processing services built on the cloud (\eg, \texttt{redhat.com}, \texttt{service-now.com}, and \texttt{appdomain.cloud}). We posit that the frequency of vulnerable subdomains in this class may be a result of bugs related to automated provisioning of servers, where latent configuration is not automatically removed during decommissioning. A second class of domains is characterized by dynamic DNS offerings and related services (\eg, \texttt{boxcast.com} and \texttt{duckdns.org}). The hosts behind these services are generally related to dynamic customers with short-lived needs, and are therefore less likely to be exploitable. We additionally saw domains that were used indirectly to host other domains (\eg, subdomains of \texttt{akadns.net}). In some cases these subdomains appear to refer to SLDs for which we also directly received traffic. The last class of domains relates to misconfigurations associated with web applications. For example, we observed a number of sessions that relate to \texttt{ring.com} (IoT doorbell cameras), \texttt{wostreaming.net} (advertising media), and \texttt{yummly.com} (recipe sharing) applications.

The right-hand side of \autoref{tbl:topdomains} shows the rankings of the discovered websites by their Tranco site ranking (those in the top 1,000 are shown). The list contains domains from entertainment, (\eg, \texttt{cnn.com}, \texttt{go.com} (Disney), and \texttt{usatoday.com}), technology, (\eg, \texttt{intel.com} and \texttt{nvidia.com}), as well as website development/hosting, (\eg, \texttt{wix.com}). Interestingly, we also saw incoming traffic to U.S. government (\eg, \texttt{hhs.gov} and \texttt{justice.gov}) services.

We also observed that the top list contains 4 academic institutions (\eg, \texttt{ harvard.edu}, \texttt{cornell.edu}, \texttt{upenn.edu}, and \texttt{jhu.edu}). This may be a reflection of the ranking system itself (which is partially based on the number of external subnets that link to the domain) that tends to favor well-known universities. These higher education institutions have large dynamic, autonomous, and fragmented web presences that may have led to server misconfigurations. In total 17 \texttt{.edu} domains exhibited latent configuration, with others under international domains such as \texttt{.edu.mx}.
\hypertarget{mr6b}{}
We additionally analyzed the subdomain depth of discovered domains. We denote subdomain depth for a given domain as the minimum depth of any discovered subdomain for that domain (\eg, \texttt{sub.example.com} has depth \SI{1}{}). Within the \SI{23}{} top \SI{1000}{} ranked domains, \SI{19}{} (\SI{83}{\percent}) had a vulnerable subdomain of depth $\le\SI{2}{}$. Among domains in the top million, the mean subdomain depth was 1.4. Among all domains, the mean subdomain depth was 0.75. This implies that some domains had dangling DNS at the SLD level---indeed, 14 top-million domains had dangling DNS records at the SLD level.

\sblock{Inferring Automation.} \hypertarget{mr6c}{}To understand what portion of latent DNS configuration was attributable to automation, we obtained two heuristic metrics: (1) \SI{325}{} (\SI{6}{\percent}) vulnerable domains had a subdomain with at least \SI{2}{} digits in it (we identify 2 digits based on conversations with affected organizations who used automation to create quasi-random subdomain names), and (2) \SI{25}{} (\SI{0.5}{\percent}) domains directly encoded the IP address in a subdomain name. While accounting for a small fraction of unique domains, digit and IP entries accounted for \SI{1159}{} (\SI{16}{\percent}) and \SI{95}{} (\SI{1.3}{\percent}) subdomains, respectively. The relatively low percentage of domains matching these criteria implies that many of the discovered vulnerabilities likely relate to manually-created domains, rather than automation.

\subsection{Disclosure and Root Causes}
\hypertarget{mr3a}{}
\label{results:root_causes}
While the majority of vulnerability disclosure was coordinated through AWS, we reached out directly to a subset of affected organizations seeking informal feedback. Because cloud service disclosures were performed through Amazon, we did not have visibility into specific affected tenants, and therefore targets for direct contact were selected based on DNS results. We contacted \SI{6}{} academic institutions, \SI{1}{} government agency, \SI{1}{} non-profit, and \SI{9}{} industrial enterprises (including \SI{6}{} high-tech, \SI{2}{} financial, and \SI{1}{} travel company), with organizations selected based on effect size, breadth, and expectation of engagement with academic research. For each of these \SI{17}{} organizations, we reached out via initial email using security contact information as available, broadly overviewing our findings and scheduling time to hold meetings with security representatives. During these scheduled meetings, we initially presented broad study results, followed by specifics of vulnerabilities found in the organization. We also outlined available data from the study that could be shared for deeper root cause analysis.

While each conference concluded with a free-form discussion of results, we primarily asked organizations to provide answers to a set of pre-scripted questions (\refappendix{appendix:questions}) after the meeting. These questions sought to understand the technical and organizational factors that led to the discovered vulnerabilities. Such qualitative results were a byproduct of our disclosure process, yet provide initial results that might motivate a more formal user study of latent configuration.

Many root causes discussed map readily to misconfiguration types discussed in prior work. Within the taxonomy introduced by Dietrich et. al~\cite{dietrich_investigating_2018}, results of the discussions generally fell under the \textit{Integration and Deployment}, \textit{No hardening}, and \textit{Scripting} type codes, with an additional \textit{Oversight} type that had not been previously considered.

\sblock{Integration \& Deployment.} The first source of latent configurations asserted by organizations was the lack of good hygiene in deploying services to the cloud. Many of these problems were the result of \textit{lift-and-shift} deployments: moving an internal service such as email or data processing from an internal server to the cloud with minimal reconfiguration or redesign. If the deployed service is not adapted to properly remove configurations upon decommissioning an instance, there is the potential for latent configuration vulnerabilities. Several of the respondents also stated that the problem was made worse because the failure is often silent. In this case, the latent configuration can exist for months or years without any indication to the affected organizations.

\sblock{No hardening.} The second source of latent configurations was attributed to simply not following best practices and established procedures (e.g., using comprehensive configuration management tools). In one case, we had an organization state that the several hosts that were identified in the study were all the result of one training organization that did not properly clean up trainee’s work. Other cases were similar. It is notable that the organizations frequently expressed that there was a need to better educate its members on best practices, and to revisit recommendations to emphasize the decommissioning phase of the service deployment life cycle.

\sblock{Scripting.} In some cases, organizations expressed that latent configurations were created as the result of automation. Rather than configuration management tools (\eg CloudFormation~\cite{aws_cloudformation} or Terraform~\cite{terraform}), these tenants had ad hoc scripts that automated creation of resources. Inevitably, the scripts did not consider the full configuration lifecycle, and ultimately introduced latent configuration vulnerabilities when the underlying compute resources were decommissioned.

\sblock{Oversight.} Underlying the above issues was the unmonitored nature of cloud use within organizations. Many organizations expressed that each part of the organization is free to create instances and configurations that are unseen by the security departments that are responsible for managing the enterprise as a whole. For example, one academic institution noted that they were aware of at least \SI{140}{} different accounts across many departments that were used to create AWS instances. This lack of control, plus the (sometimes) lack of sophistication by the users provisioning cloud resources, led to some organizations seeing cloud use as a kind of technical ``wild west''.

\sblock{Ethics.} \hypertarget{mr1b}{}Our disclosure process (\refappendix{appendix:disclosure}) ensured that no personal information would be conveyed during disclosure, as all contacted parties were representatives of organizations and not speaking as individuals. As such, our disclosures were determined to be exempt by our institutional review board (IRB) as not human subjects research.

\section{Defenses and Mitigations}
\label{sec:mitigations}

Cloud Squatting ultimately results from a tenant's failure to properly decommission configuration. As a result, the most compelling defenses take the form of best practices by tenants. Even when tenants are unable to adopt best practices, however, cloud providers can take actions to reduce the prevalence and exploitability of latent configuration. We begin by surveying known defenses against dangling DNS that generalize to latent configuration, then discuss the effect of IP allocation policies on mitigating the risk of IP reuse. We conclude by discussing the actions taken by Amazon in response to our study.

\subsection{Best Practices}\hypertarget{mr4a}{}

\sblock{Preventing misconfiguration.} Classical web service models often relied on static IPs assigned to services~\cite{noauthor_best_2019}, with long-lived configuration such as DNS referencing these addresses. In public clouds, this assumption can break down to the detriment of security. When designing services on public clouds, care should be taken to ensure that references to service IPs are either managed by the cloud provider (\ie, the cloud provider resolves a unique resource, such as a domain name, to an IP address while ensuring latent configuration is prevented), or some configuration manager or policy. In each case, cloud providers also have the opportunity to encourage best practices and alert tenants when latent configuration exists.

\sblock{Leveraging DNS.} While our work focuses on configuration that exists beyond DNS, many in the community have demonstrated compelling defenses when DNS is used as part of configuration. For instance, recent work suggests that changes to TLS certificate issuance can be made resistant to dangling DNS~\cite{borgolte_cloud_2018}, and using such securely-allocated certificates for server/client authentication would prevent latent configuration from being exploited. As such, configurations that do not use DNS (\eg raw IP addresses or IPBNs configured as cloud service endpoints) can be reconfigured to resolve to the address through DNS. In this way, existing solutions toward preventing dangling DNS can be applied to these configurations.

DNS can also serve as a repository for automated configuration management. When provisioning instances with public IP addresses on AWS, tenants are currently provided with an IP-based DNS name (IPBN). Cloud providers could replace such IP-based names with names that use unique identifiers, such as the instance ID. When the service is deprovisioned, such DNS records would be deleted automatically, thereby preventing latent configuration. While Amazon recently released such a feature (Resource-based naming~\cite{noauthor_introducing_2021}) for private IP addresses, it is at present not supported for public IPs.

\sblock{TLS and pre-shared keys.} When client and server can authenticate using keys created independently of latent configuration, the latent configuration no longer poses a threat. For instance, applications could contain a pre-shared certificate for server communication or require a certificate authority that does not validate certificates based on DNS. This practice is often referred to as leaf or CA certificate pinning~\cite{oltrogge_pin_2015}.

\sblock{Preventing IP Reuse.} IP reuse is a unique property of public clouds using shared IP pools, and preventing this reuse ensures that services cannot fall victim to IP use-after-free exploits. This can be accomplished through existing bring your own IP (BYOIP) offerings, which allow tenants to migrate their owned IPv4 ranges to a public cloud. Services provisioned within this IP space will not have their IP addresses reused by other tenants, preventing cloud squatting. Cloud providers can improve best practices and documentation to demonstrate this benefit, and encourage larger tenants to leverage BYOIP to benefit security.\hypertarget{or3a}{} Transitioning to IPv6 for networking also prevents IP reuse, as the large address space can be segmented per tenant. Unfortunately, publicly-facing services (\eg, logging, metrics, webhooks, and API endpoints) must still generally be exposed at some IPv4 endpoint for compatibility~\cite{noauthor_ipv6_nodate}.

In other cases, using public IP addresses is unnecessary. For instance, some managed cloud services (such as Application Gateway) can communicate with backend services through private IP addresses (known within AWS as Virtual Private Cloud or VPC~\cite{amazon_vpc}). When private IP addresses are used, there is no risk of reuse by other tenants, as the IP address is only valid within the context of the individual tenant. When this route is available, cloud squatting is effectively prevented.

\subsection{Mitigations}

\sblock{Detecting latent configuration.}
When a cloud provider has a complete view of configuration and IP address allocation (\ie, managed cloud services connecting to cloud-managed IPs), they can detect when a tenant references an IP address that they no longer control. Control-plane information from a given cloud service can be cross-referenced against IP allocation logs and alerts can be automatically sent to tenants. In addition to cross-referencing configuration directly referencing IP addresses, cloud providers could perform DNS resolution on domains used in configuration to determine if they reference a tenant's previously-controlled IP address.

Dangling DNS records are a clear place to start. \texttt{A} records to raw AWS IPs or \texttt{CNAME} records that resolve to raw IPs (\eg, via IPBNs) can lead to a vulnerability when the instance is decommissioned but DNS records are not removed. When DNS resolution and IP allocation are both controlled by the cloud provider, remediation is possible. \texttt{A} records configured through the provider's DNS can be cross-checked against IP allocation without interacting with any tenant resources. In addition, the cloud provider might be able to remove a DNS record automatically. More broadly, cloud providers can play a key role in preventing latent DNS configuration by discouraging the use of raw IP/IPBN references.

The ability to cross-reference control-plane information also opens the possibility for cloud providers to interactively notify users of latent configuration during decommissioning. When a user decommissions a server, services within the same account could be checked for references to the IP in real-time, with users given the option to directly remove latent configuration along with the cloud server. For example, such a check would prevent the leakage of data through SNS traffic when servers are improperly decommissioned.
While such an approach may require additional complexity in cloud management consoles, it may dramatically reduce the incidence of latent configuration for manually-managed services.

\sblock{IP Allocation Policy.}
Cloud providers currently allocate IPs pseudo-randomly from a pool. New allocation policies can prevent adversaries from exploiting a large number of IP addresses, while being transparent to tenants and complementary to other defenses. We propose \textit{IP Tagging}, and evaluate it against the existing random allocation, and a least-recently-used (LRU, the oldest address is always allocated) allocation.

Under IP Tagging, when an IP address is released, it is tagged with both the release time and the tenant. When a new IP is requested, preference is first given to IP addresses that the tenant previously released, followed by the address that has been in the pool the longest. Tagging prevents cloud squatting in multiple ways: (1) Adversaries are prevented from scanning the entire IP space by allocating many instances, (2) tenants receive their same IP addresses back, reducing the number of tenants associated with each IP address and therefore the likelihood of any individual IP address being exploitable---in essence allowing the allocations to self-partition by tenant.

We perform a brief experiment simulating policies on a cloud IP address pool. Simulated \textit{tenant agents} allocate and deallocate IPs from the pool randomly up to a quota. An \textit{adversarial agent} allocates IP addresses with the goal of observing traffic intended for previous tenants. IPs are allocated for a fixed duration with a maximum quota of simultaneous IPs allocated (as observed in AWS). The simulation measures the efficacy of each allocation policy through (a) the number of unique addresses allocated to the simulated adversary, (b) how long ago a previous tenant controlled the assigned addresses, and (c) how many tenants are associated with each address.

Our simulation matches parameters observed in \texttt{us-east-1a}, with \SI{673}{k} unique IPs modeled. \SI{100}{k} tenants are modeled, with a new tenant quota selected at random every \SI{10}{} minutes. IPs are allocated and deallocated to reach this target and assigned to tenants at random. The simulator could also be augmented with actual IP allocation traces from a public cloud, though such a dataset is not available. The adversarial agent holds IP addresses for \SI{10}{} minutes with a quota of \SI{60}{} addresses (similar to the per-zone quota from our measurement study). A total of \SI{581}{k} IPs were allocated by the adversary under each policy, again mirroring our actual experiment.

\begin{sidebar}
\section*{Amazon Actions}

In response to this study and a subsequent internal audit of AWS deployments, Amazon is performing the following actions to assist AWS customers: 

\vspace{3pt}
\noindent
\textbf{Cloud Configuration.} When cloud services can be configured to interact with an AWS compute resource, the management console is being updated to alert users when they subscribe elastic IP addresses directly to SNS Topics or health checks. 

\vspace{3pt}
\noindent
\textbf{Expanded scanning/disclosure of vulnerabilities.} Amazon is developing tools that analyze control-plane information to locate customers with current misconfigurations across all tenants and regions. The outputs of scans will be used to send notices to customers with misconfigured cloud services to review their configuration for SNS topics and Route53 heath checks.

\vspace{3pt}
\noindent
\textbf{Automated Policy Enforcement.} Amazon is developing managed Config rules that customers can apply to their accounts  within an organization. These Config Rules can be configured to prevent, remediate, or alert on cloud assets that meet the conditions of the rules. 

\vspace{3pt}
\noindent
\textbf{Updated Best Practices.} AWS is updating Route53 and SNS best practices documentation to recommend customers avoid tying services and configurations to elastic IPs and ensuring good hygiene for server instantiating and decommissioning. For instance, best practice documentation for SNS will discuss the risks of failing to remove subscriptions, especially when raw IP addresses of AWS instances and unencrypted messages are used.

\end{sidebar}

\begin{table}[]
    \caption{Experimental results of IP pool simulation.}
    \centering
    \begin{tabular}{cccc}
        \toprule
        Policy & Unique IPs & Mean Prev. Tenants & Median Reuse Time \\
        \midrule
        \textsc{Random} & \SI{377596}{} & \SI{228.2}{} & \SI{5.7E3}{s} \\
        \textsc{LRU} & \SI{385774}{} & \SI{209.6}{} & \SI{9.2E3}{s}\\
        \textsc{Tagging} & \SI{240}{} & \SI{2.387}{} & \SI{2.9E6}{s}\\
        \bottomrule
    \end{tabular}
    \label{tab:tagging_results}
\end{table}

Our results are shown in \autoref{tab:tagging_results}. While an LRU policy increases the median reuse time (\SI{62}{\percent} increase vs random allocation), IP Tagging has the greatest impact on all three metrics. Unique IPs are reduced by \SI{99.94}{\percent}, mean previous tenants by \SI{98.95}{\percent}, and reuse time is increased by \SI{514}{x} compared to random allocation. Note that, while allocating more addresses under both LRU and \textsc{Random} would further increase coverage of unique IPs, under the Tagging policy IPs have been limited to a constant: with \SI{60}{} simultaneous IPs allocated for \SI{10}{} minutes each and a cooldown of \SI{30}{min} only \SI{240}{} IPs are ever seen.

Our analysis of IP tagging currently assumes a single-tenant adversary, wherein the cloud provider can apply the policy to effectively prevent a large number of IPs from being observed. An adversary could partially bypass this mitigation using multiple cloud accounts. In this case, however, IP tagging still has benefits: (1) reduced overall contention in the address pool would increase the interval between IP address reuse, providing more time for latent configuration to be removed, and (2) previously-used IPs by benign tenants would be recycled to the tenant, reducing the number of IPs with which they could potentially associate latent configuration. Further investigation on the affects of IP allocation policies in clouds is warranted in future works.

\subsection{Deployed Mitigations}\label{deployed_mitigations}

In the interest of ethical disclosure (see \refappendix{appendix:disclosure}) we worked directly with AWS to share results and discuss mitigations. Our initial discussions began in June 2021 and continue as of the submission of this paper. AWS has performed internal reviews which confirmed potential misconfigurations existed over all AWS regions. In response and detailed in the side-bar (reviewed and confirmed to be accurate by AWS on August 17, 2021), they have taken a number of actions to help their customer community. Several of these actions are similar to those we identified, but others are oriented toward existing AWS services (and thus are less general-purpose solutions). However, those, like ours are directed towards many of the same root cause issues---technical and organizational---that we identified in the study and disclosure.

\section{Conclusions}\label{conclusion}

The advantages of public clouds are not without architectural and security risks: our study has confirmed that, not only are latent configuration vulnerabilities prevalent across organizations of all sizes and verticals, but the classes of configurations leading to these vulnerabilities are diverse. We conclude that latent configuration represents a fundamental security risk in the shared networking environment of public clouds, and emphasize that care must be taken by cloud tenants such that service decommissioning does not introduce latent configuration. At the same time, our investigation of mitigations shows that there is reason for optimism: scanning techniques available to---and soon to be deployed by---cloud providers can detect and potentially correct latent configuration vulnerabilities automatically, and changes to IP pool allocation prevent exploitation of vulnerabilities.

Root causes of latent configuration span beyond purely technical. Affected parties overwhelmingly emphasized organizational concerns: large organizations using public clouds often have multiple accounts (up to \SI{140}{} accounts in one case) provisioning services with minimal central oversight. Further complicating cloud use for organizations is the emphasis on transitioning on-premises services to the cloud (lift-and-shift), which additionally transfers the implicit assumptions of these services such as non-shared resources. Because of the shared responsibility model, tenants will face the consequences of latent configuration unless it is properly managed. 

Beyond the mitigations presented, our work suggests that latent configuration is a fundamental risk of shared computing environments such as public clouds. However, cloud providers and the security community generally have the opportunity to improve the security postures of cloud tenants: existing and new best practices can reduce dependence on configurations that may become latent. Automation also plays a key role: platforms that are sensitive to configuration hygiene can be secure by default, providing guarantees against latent configuration. Cloud providers can further assist by providing policy enforcement of safe configuration at the organization-level. In these ways, the benefits of public clouds can be enjoyed while also ensuring security of tenant services.
\bibliographystyle{IEEEtran}
\bibliography{references}

% Generated by IEEEtran.bst, version: 1.14 (2015/08/26)
\begin{thebibliography}{10}
\providecommand{\url}[1]{#1}
\csname url@samestyle\endcsname
\providecommand{\newblock}{\relax}
\providecommand{\bibinfo}[2]{#2}
\providecommand{\BIBentrySTDinterwordspacing}{\spaceskip=0pt\relax}
\providecommand{\BIBentryALTinterwordstretchfactor}{4}
\providecommand{\BIBentryALTinterwordspacing}{\spaceskip=\fontdimen2\font plus
\BIBentryALTinterwordstretchfactor\fontdimen3\font minus
  \fontdimen4\font\relax}
\providecommand{\BIBforeignlanguage}[2]{{%
\expandafter\ifx\csname l@#1\endcsname\relax
\typeout{** WARNING: IEEEtran.bst: No hyphenation pattern has been}%
\typeout{** loaded for the language `#1'. Using the pattern for}%
\typeout{** the default language instead.}%
\else
\language=\csname l@#1\endcsname
\fi
#2}}
\providecommand{\BIBdecl}{\relax}
\BIBdecl

\bibitem{aws_website}
\BIBentryALTinterwordspacing
``\BIBforeignlanguage{en-US}{Cloud {Services} - {Amazon} {Web} {Services}
  ({AWS})}.'' [Online]. Available: \url{https://aws.amazon.com/}
\BIBentrySTDinterwordspacing

\bibitem{google_cloud_website}
\BIBentryALTinterwordspacing
``Cloud {Computing} {Services}  {\textbar}  {Google} {Cloud}.'' [Online].
  Available: \url{https://cloud.google.com/}
\BIBentrySTDinterwordspacing

\bibitem{azure_website}
\BIBentryALTinterwordspacing
``\BIBforeignlanguage{en}{Cloud {Computing} {Services} {\textbar} {Microsoft}
  {Azure}}.'' [Online]. Available: \url{https://azure.microsoft.com/en-us/}
\BIBentrySTDinterwordspacing

\bibitem{borgolte_cloud_2018}
\BIBentryALTinterwordspacing
K.~Borgolte, T.~Fiebig, S.~Hao, C.~Kruegel, and G.~Vigna,
  ``\BIBforeignlanguage{en}{Cloud {Strife}: {Mitigating} the {Security} {Risks}
  of {Domain}-{Validated} {Certificates}},'' in
  \emph{\BIBforeignlanguage{en}{Proceedings 2018 {Network} and {Distributed}
  {System} {Security} {Symposium}}}.\hskip 1em plus 0.5em minus 0.4em\relax San
  Diego, CA: Internet Society, 2018. [Online]. Available:
  \url{https://www.ndss-symposium.org/wp-content/uploads/2018/02/ndss2018_06A-4_Borgolte_paper.pdf}
\BIBentrySTDinterwordspacing

\bibitem{liu_all_2016}
\BIBentryALTinterwordspacing
D.~Liu, S.~Hao, and H.~Wang, ``\BIBforeignlanguage{en}{All {Your} {DNS}
  {Records} {Point} to {Us}: {Understanding} the {Security} {Threats} of
  {Dangling} {DNS} {Records}},'' in \emph{\BIBforeignlanguage{en}{Proceedings
  of the 2016 {ACM} {SIGSAC} {Conference} on {Computer} and {Communications}
  {Security}}}.\hskip 1em plus 0.5em minus 0.4em\relax Vienna Austria: ACM,
  Oct. 2016, pp. 1414--1425. [Online]. Available:
  \url{https://dl.acm.org/doi/10.1145/2976749.2978387}
\BIBentrySTDinterwordspacing

\bibitem{sns}
\BIBentryALTinterwordspacing
``\BIBforeignlanguage{en-US}{Amazon {Simple} {Notification} {Service} ({SNS})
  {\textbar} {Messaging} {Service} {\textbar} {AWS}}.'' [Online]. Available:
  \url{https://aws.amazon.com/sns/}
\BIBentrySTDinterwordspacing

\bibitem{route53}
\BIBentryALTinterwordspacing
``\BIBforeignlanguage{en-US}{Amazon {Route} 53 - {Amazon} {Web} {Services}}.''
  [Online]. Available: \url{https://aws.amazon.com/route53/}
\BIBentrySTDinterwordspacing

\bibitem{cloudfront}
\BIBentryALTinterwordspacing
``\BIBforeignlanguage{en-US}{Content {Delivery} {Network} ({CDN}) {\textbar}
  {Low} {Latency}, {High} {Transfer} {Speeds}, {Video} {Streaming} {\textbar}
  {Amazon} {CloudFront}}.'' [Online]. Available:
  \url{https://aws.amazon.com/cloudfront/}
\BIBentrySTDinterwordspacing

\bibitem{api_gateway}
\BIBentryALTinterwordspacing
``\BIBforeignlanguage{en-US}{Amazon {API} {Gateway} {\textbar} {API}
  {Management} {\textbar} {Amazon} {Web} {Services}}.'' [Online]. Available:
  \url{https://aws.amazon.com/api-gateway/}
\BIBentrySTDinterwordspacing

\bibitem{postgres}
\BIBentryALTinterwordspacing
P.~G.~D. Group, ``\BIBforeignlanguage{en}{{PostgreSQL}},'' Aug. 2021. [Online].
  Available: \url{https://www.postgresql.org/}
\BIBentrySTDinterwordspacing

\bibitem{elasticsearch}
\BIBentryALTinterwordspacing
``\BIBforeignlanguage{en-us}{Free and {Open} {Search}: {The} {Creators} of
  {Elasticsearch}, {ELK} \& {Kibana} {\textbar} {Elastic}}.'' [Online].
  Available: \url{https://www.elastic.co/}
\BIBentrySTDinterwordspacing

\bibitem{mysql}
\BIBentryALTinterwordspacing
``{MySQL}.'' [Online]. Available: \url{https://www.mysql.com/}
\BIBentrySTDinterwordspacing

\bibitem{redis}
\BIBentryALTinterwordspacing
``Redis.'' [Online]. Available: \url{https://redis.io/}
\BIBentrySTDinterwordspacing

\bibitem{fix_protocol}
\BIBentryALTinterwordspacing
F.~T. Community, ``\BIBforeignlanguage{en-US}{{FIX} {Latest} {Online}
  {Specification} • {FIX} {Trading} {Community}}.'' [Online]. Available:
  \url{https://www.fixtrading.org/online-specification/}
\BIBentrySTDinterwordspacing

\bibitem{teamwork}
\BIBentryALTinterwordspacing
``Teamwork: {The} {Best} {Platform} for {Client} {Work}.'' [Online]. Available:
  \url{https://www.teamwork.com/}
\BIBentrySTDinterwordspacing

\bibitem{bitbucket}
\BIBentryALTinterwordspacing
Atlassian, ``\BIBforeignlanguage{en}{Bitbucket {\textbar} {The} {Git} solution
  for professional teams}.'' [Online]. Available:
  \url{https://bitbucket.org/product}
\BIBentrySTDinterwordspacing

\bibitem{segment}
\BIBentryALTinterwordspacing
``\BIBforeignlanguage{en}{Segment {\textbar} \#1 {CDP} to {Manage} {Customer}
  {Data}}.'' [Online]. Available: \url{https://segment.com/}
\BIBentrySTDinterwordspacing

\bibitem{antonakakis_understanding_nodate}
\BIBentryALTinterwordspacing
M.~Antonakakis, T.~April, M.~Bailey, M.~Bernhard, E.~Bursztein, J.~Cochran,
  Z.~Durumeric, J.~A. Halderman, L.~Invernizzi, M.~Kallitsis, D.~Kumar,
  C.~Lever, Z.~Ma, J.~Mason, D.~Menscher, C.~Seaman, N.~Sullivan, K.~Thomas,
  and Y.~Zhou, ``Understanding the mirai botnet,'' in \emph{26th {USENIX}
  Security Symposium ({USENIX} Security 17)}.\hskip 1em plus 0.5em minus
  0.4em\relax Vancouver, BC: {USENIX} Association, Aug. 2017, pp. 1093--1110.
  [Online]. Available:
  \url{https://www.usenix.org/conference/usenixsecurity17/technical-sessions/presentation/antonakakis}
\BIBentrySTDinterwordspacing

\bibitem{durumeric_search_2015}
\BIBentryALTinterwordspacing
Z.~Durumeric, D.~Adrian, A.~Mirian, M.~Bailey, and J.~A. Halderman,
  ``\BIBforeignlanguage{en}{A {Search} {Engine} {Backed} by {Internet}-{Wide}
  {Scanning}},'' in \emph{\BIBforeignlanguage{en}{Proceedings of the 22nd {ACM}
  {SIGSAC} {Conference} on {Computer} and {Communications} {Security}}}.\hskip
  1em plus 0.5em minus 0.4em\relax Denver Colorado USA: ACM, Oct. 2015, pp.
  542--553. [Online]. Available:
  \url{https://dl.acm.org/doi/10.1145/2810103.2813703}
\BIBentrySTDinterwordspacing

\bibitem{wustrow_internet_2010}
\BIBentryALTinterwordspacing
E.~Wustrow, M.~Karir, M.~Bailey, F.~Jahanian, and G.~Huston,
  ``\BIBforeignlanguage{en}{Internet background radiation revisited},'' in
  \emph{\BIBforeignlanguage{en}{Proceedings of the 10th annual conference on
  {Internet} measurement - {IMC} '10}}.\hskip 1em plus 0.5em minus 0.4em\relax
  Melbourne, Australia: ACM Press, 2010, p.~62. [Online]. Available:
  \url{http://portal.acm.org/citation.cfm?doid=1879141.1879149}
\BIBentrySTDinterwordspacing

\bibitem{bailey_internet_2005}
M.~Bailey, E.~Cooke, F.~Jahanian, J.~Nazario, and D.~Watson, ``The internet
  motion sensor-a distributed blackhole monitoring system.'' in
  \emph{{NDSS}}.\hskip 1em plus 0.5em minus 0.4em\relax Citeseer, 2005.

\bibitem{protalinski_microsoft_2010}
\BIBentryALTinterwordspacing
E.~Protalinski, ``\BIBforeignlanguage{en-us}{Microsoft repackages its
  productivity services as {Office} 365},'' Oct. 2010. [Online]. Available:
  \url{https://arstechnica.com/microsoft/news/2010/10/microsoft-repackages-its-productivity-services-as-office-365.ars}
\BIBentrySTDinterwordspacing

\bibitem{geuss_ibm_2016}
\BIBentryALTinterwordspacing
M.~Geuss, ``\BIBforeignlanguage{en-us}{{IBM} wants to move blockchain tech
  beyond {Bitcoin} and money transfer},'' Feb. 2016. [Online]. Available:
  \url{https://arstechnica.com/information-technology/2016/02/ibm-wants-to-move-blockchain-tech-beyond-bitcoin-and-money-transfer/}
\BIBentrySTDinterwordspacing

\bibitem{wang2010impact}
G.~Wang and T.~E. Ng, ``The impact of virtualization on network performance of
  amazon ec2 data center,'' in \emph{2010 Proceedings IEEE INFOCOM}.\hskip 1em
  plus 0.5em minus 0.4em\relax IEEE, 2010, pp. 1--9.

\bibitem{herbst2013elasticity}
N.~R. Herbst, S.~Kounev, and R.~Reussner, ``Elasticity in cloud computing: What
  it is, and what it is not,'' in \emph{10th international conference on
  autonomic computing (ICAC 13)}, 2013, pp. 23--27.

\bibitem{dillon_cloud_2010}
T.~Dillon, C.~Wu, and E.~Chang, ``Cloud {Computing}: {Issues} and
  {Challenges},'' in \emph{2010 24th {IEEE} {International} {Conference} on
  {Advanced} {Information} {Networking} and {Applications}}, Apr. 2010, pp.
  27--33, iSSN: 2332-5658.

\bibitem{xu_early_nodate}
\BIBentryALTinterwordspacing
T.~Xu, X.~Jin, P.~Huang, Y.~Zhou, S.~Lu, L.~Jin, and S.~Pasupathy, ``Early
  detection of configuration errors to reduce failure damage,'' in \emph{12th
  {USENIX} Symposium on Operating Systems Design and Implementation ({OSDI}
  16)}.\hskip 1em plus 0.5em minus 0.4em\relax Savannah, GA: {USENIX}
  Association, Nov. 2016, pp. 619--634. [Online]. Available:
  \url{https://www.usenix.org/conference/osdi16/technical-sessions/presentation/xu}
\BIBentrySTDinterwordspacing

\bibitem{alowaisheq_zombie_2020}
\BIBentryALTinterwordspacing
E.~Alowaisheq, S.~Tang, Z.~Wang, F.~Alharbi, X.~Liao, and X.~Wang,
  ``\BIBforeignlanguage{en}{Zombie {Awakening}: {Stealthy} {Hijacking} of
  {Active} {Domains} through {DNS} {Hosting} {Referral}},'' in
  \emph{\BIBforeignlanguage{en}{Proceedings of the 2020 {ACM} {SIGSAC}
  {Conference} on {Computer} and {Communications} {Security}}}.\hskip 1em plus
  0.5em minus 0.4em\relax Virtual Event USA: ACM, Oct. 2020, pp. 1307--1322.
  [Online]. Available: \url{https://dl.acm.org/doi/10.1145/3372297.3417864}
\BIBentrySTDinterwordspacing

\bibitem{noauthor_dangling_2021}
\BIBentryALTinterwordspacing
``\BIBforeignlanguage{en-US}{Dangling {Domains}: {Security} {Threats},
  {Detection} and {Prevalence}},'' Sep. 2021. [Online]. Available:
  \url{https://unit42.paloaltonetworks.com/dangling-domains/}
\BIBentrySTDinterwordspacing

\bibitem{noauthor_lets_nodate}
\BIBentryALTinterwordspacing
``Let's {Encrypt} - {Free} {SSL}/{TLS} {Certificates}.'' [Online]. Available:
  \url{https://letsencrypt.org/}
\BIBentrySTDinterwordspacing

\bibitem{kintis_hiding_2017}
\BIBentryALTinterwordspacing
P.~Kintis, N.~Miramirkhani, C.~Lever, Y.~Chen, R.~Romero-Gómez, N.~Pitropakis,
  N.~Nikiforakis, and M.~Antonakakis, ``Hiding in {Plain} {Sight}: {A}
  {Longitudinal} {Study} of {Combosquatting} {Abuse},'' in \emph{Proceedings of
  the 2017 {ACM} {SIGSAC} {Conference} on {Computer} and {Communications}
  {Security}}, ser. {CCS} '17.\hskip 1em plus 0.5em minus 0.4em\relax New York,
  NY, USA: Association for Computing Machinery, Oct. 2017, pp. 569--586.
  [Online]. Available: \url{https://doi.org/10.1145/3133956.3134002}
\BIBentrySTDinterwordspacing

\bibitem{szurdi_long_2014}
J.~Szurdi, B.~Kocso, G.~Cseh, J.~Spring, M.~Felegyhazi, and C.~Kanich, ``The
  long “taile” of typosquatting domain names,'' in \emph{23rd {USENIX}
  {Security} {Symposium} ({USENIX}{Security} 14)}, 2014, pp. 191--206.

\bibitem{vijayakumar_sting_2012}
H.~Vijayakumar, J.~Schiffman, and T.~Jaeger, ``{STING}: {Finding} {Name}
  {Resolution} {Vulnerabilities} in {Programs},'' in \emph{21st {USENIX}
  {Security} {Symposium} ({USENIX} {Security} 12)}, 2012, pp. 585--599.

\bibitem{kumar_skill_2018}
\BIBentryALTinterwordspacing
D.~Kumar, R.~Paccagnella, P.~Murley, E.~Hennenfent, J.~Mason, A.~Bates, and
  M.~Bailey, ``\BIBforeignlanguage{en}{Skill {Squatting} {Attacks} on {Amazon}
  {Alexa}},'' 2018, pp. 33--47. [Online]. Available:
  \url{https://www.usenix.org/conference/usenixsecurity18/presentation/kumar}
\BIBentrySTDinterwordspacing

\bibitem{noauthor_building_2017}
\BIBentryALTinterwordspacing
``\BIBforeignlanguage{en-US}{Building {Scalable} {Applications} and
  {Microservices}: {Adding} {Messaging} to {Your} {Toolbox}},'' May 2017,
  section: Amazon EC2. [Online]. Available:
  \url{https://aws.amazon.com/blogs/compute/building-scalable-applications-and-microservices-adding-messaging-to-your-toolbox/}
\BIBentrySTDinterwordspacing

\bibitem{mqtt}
\BIBentryALTinterwordspacing
``{MQTT} - {The} {Standard} for {IoT} {Messaging}.'' [Online]. Available:
  \url{https://mqtt.org/}
\BIBentrySTDinterwordspacing

\bibitem{aws_dns_hostnames}
\BIBentryALTinterwordspacing
``{DNS support for your VPC}.'' [Online]. Available:
  \url{https://docs.aws.amazon.com/vpc/latest/userguide/vpc-dns.html}
\BIBentrySTDinterwordspacing

\bibitem{noauthor_amazon_2021}
\BIBentryALTinterwordspacing
``\BIBforeignlanguage{en}{Amazon and {Google} patch major bug in their
  {DNS}-as-a-{Service} platforms},'' Aug. 2021. [Online]. Available:
  \url{https://therecord.media/amazon-and-google-patch-major-bug-in-their-dns-as-a-service-platforms/}
\BIBentrySTDinterwordspacing

\bibitem{paz_t-mobile_2021}
\BIBentryALTinterwordspacing
I.~G. Paz, ``\BIBforeignlanguage{en-US}{T-{Mobile} {Says} {Hack} {Exposed}
  {Personal} {Data} of 40 {Million} {People}},''
  \emph{\BIBforeignlanguage{en-US}{The New York Times}}, Aug. 2021. [Online].
  Available:
  \url{https://www.nytimes.com/2021/08/18/business/tmobile-data-breach.html}
\BIBentrySTDinterwordspacing

\bibitem{dhamija_why_2006}
\BIBentryALTinterwordspacing
R.~Dhamija, J.~D. Tygar, and M.~Hearst, ``Why phishing works,'' in
  \emph{Proceedings of the {SIGCHI} {Conference} on {Human} {Factors} in
  {Computing} {Systems}}.\hskip 1em plus 0.5em minus 0.4em\relax New York, NY,
  USA: Association for Computing Machinery, Apr. 2006, pp. 581--590. [Online].
  Available: \url{https://doi.org/10.1145/1124772.1124861}
\BIBentrySTDinterwordspacing

\bibitem{hong_state_2012}
\BIBentryALTinterwordspacing
J.~Hong, ``The state of phishing attacks,'' \emph{Communications of the ACM},
  vol.~55, no.~1, pp. 74--81, Jan. 2012. [Online]. Available:
  \url{https://doi.org/10.1145/2063176.2063197}
\BIBentrySTDinterwordspacing

\bibitem{kharraz_cutting_2015}
A.~Kharraz, W.~Robertson, D.~Balzarotti, L.~Bilge, and E.~Kirda,
  ``\BIBforeignlanguage{en}{Cutting the {Gordian} {Knot}: {A} {Look} {Under}
  the {Hood} of {Ransomware} {Attacks}},'' in
  \emph{\BIBforeignlanguage{en}{Detection of {Intrusions} and {Malware}, and
  {Vulnerability} {Assessment}}}, ser. Lecture {Notes} in {Computer} {Science},
  M.~Almgren, V.~Gulisano, and F.~Maggi, Eds.\hskip 1em plus 0.5em minus
  0.4em\relax Cham: Springer International Publishing, 2015, pp. 3--24.

\bibitem{laszka_economics_2017}
A.~Laszka, S.~Farhang, and J.~Grossklags, ``\BIBforeignlanguage{en}{On the
  {Economics} of {Ransomware}},'' in \emph{\BIBforeignlanguage{en}{Decision and
  {Game} {Theory} for {Security}}}, ser. Lecture {Notes} in {Computer}
  {Science}, S.~Rass, B.~An, C.~Kiekintveld, F.~Fang, and S.~Schauer,
  Eds.\hskip 1em plus 0.5em minus 0.4em\relax Cham: Springer International
  Publishing, 2017, pp. 397--417.

\bibitem{al-bataineh_analysis_2012}
A.~Al-Bataineh and G.~White, ``Analysis and detection of malicious data
  exfiltration in web traffic,'' in \emph{2012 7th {International} {Conference}
  on {Malicious} and {Unwanted} {Software}}, Oct. 2012, pp. 26--31.

\bibitem{ullah_data_2018}
\BIBentryALTinterwordspacing
F.~Ullah, M.~Edwards, R.~Ramdhany, R.~Chitchyan, M.~A. Babar, and A.~Rashid,
  ``\BIBforeignlanguage{en}{Data exfiltration: {A} review of external attack
  vectors and countermeasures},'' \emph{\BIBforeignlanguage{en}{Journal of
  Network and Computer Applications}}, vol. 101, pp. 18--54, Jan. 2018.
  [Online]. Available:
  \url{https://www.sciencedirect.com/science/article/pii/S1084804517303569}
\BIBentrySTDinterwordspacing

\bibitem{antonakakis_understanding_2017}
M.~Antonakakis, T.~April, M.~Bailey, M.~Bernhard, E.~Bursztein, J.~Cochran,
  Z.~Durumeric, J.~A. Halderman, L.~Invernizzi, and M.~Kallitsis,
  ``Understanding the mirai botnet,'' in \emph{26th {USENIX} security symposium
  ({USENIX} {Security} 17)}, 2017, pp. 1093--1110.

\bibitem{pang_characteristics_2004}
\BIBentryALTinterwordspacing
R.~Pang, V.~Yegneswaran, P.~Barford, V.~Paxson, and L.~Peterson,
  ``\BIBforeignlanguage{en}{Characteristics of internet background
  radiation},'' in \emph{\BIBforeignlanguage{en}{Proceedings of the 4th {ACM}
  {SIGCOMM} conference on {Internet} measurement - {IMC} '04}}.\hskip 1em plus
  0.5em minus 0.4em\relax Taormina, Sicily, Italy: ACM Press, 2004, p.~27.
  [Online]. Available:
  \url{http://portal.acm.org/citation.cfm?doid=1028788.1028794}
\BIBentrySTDinterwordspacing

\bibitem{richter_scanning_2019}
\BIBentryALTinterwordspacing
P.~Richter and A.~Berger, ``Scanning the {Scanners}: {Sensing} the {Internet}
  from a {Massively} {Distributed} {Network} {Telescope},'' in
  \emph{Proceedings of the {Internet} {Measurement} {Conference}}, ser. {IMC}
  '19.\hskip 1em plus 0.5em minus 0.4em\relax New York, NY, USA: Association
  for Computing Machinery, Oct. 2019, pp. 144--157. [Online]. Available:
  \url{https://doi.org/10.1145/3355369.3355595}
\BIBentrySTDinterwordspacing

\bibitem{sandland_statistical_1984}
\BIBentryALTinterwordspacing
R.~L. Sandland and R.~M. Cormack, ``Statistical {Inference} for {Poisson} and
  {Multinomial} {Models} for {Capture}- {Recapture} {Experiments},''
  \emph{Biometrika}, vol.~71, no.~1, pp. 27--33, 1984, publisher: [Oxford
  University Press, Biometrika Trust]. [Online]. Available:
  \url{https://www.jstor.org/stable/2336393}
\BIBentrySTDinterwordspacing

\bibitem{baillargeon_rcapture_2007}
\BIBentryALTinterwordspacing
S.~Baillargeon and L.-P. Rivest, ``\BIBforeignlanguage{en}{Rcapture:
  {Loglinear} {Models} for {Capture}-{Recapture} in {R}},''
  \emph{\BIBforeignlanguage{en}{Journal of Statistical Software}}, vol.~19,
  no.~1, pp. 1--31, Apr. 2007, number: 1. [Online]. Available:
  \url{https://www.jstatsoft.org/index.php/jss/article/view/v019i05}
\BIBentrySTDinterwordspacing

\bibitem{aws_ip_ranges}
\BIBentryALTinterwordspacing
``{AWS} {IP} address ranges - {AWS} {General} {Reference}.'' [Online].
  Available:
  \url{https://docs.aws.amazon.com/general/latest/gr/aws-ip-ranges.html}
\BIBentrySTDinterwordspacing

\bibitem{azure_event_grid}
\BIBentryALTinterwordspacing
``\BIBforeignlanguage{en-us}{{WebHook} event delivery - {Azure} {Event}
  {Grid}}.'' [Online]. Available:
  \url{https://docs.microsoft.com/en-us/azure/event-grid/webhook-event-delivery}
\BIBentrySTDinterwordspacing

\bibitem{gcp_pubsub}
\BIBentryALTinterwordspacing
``\BIBforeignlanguage{en}{Push subscriptions {\textbar} {Cloud} {Pub}/{Sub}
  {Documentation}}.'' [Online]. Available:
  \url{https://cloud.google.com/pubsub/docs/push}
\BIBentrySTDinterwordspacing

\bibitem{tsaousis_firehol_nodate}
\BIBentryALTinterwordspacing
C.~Tsaousis, ``\BIBforeignlanguage{en}{{FireHOL} {IP} {Lists} {\textbar} {IP}
  {Blacklists} {\textbar} {IP} {Reputation} {Feeds}}.'' [Online]. Available:
  \url{http://iplists.firehol.org/}
\BIBentrySTDinterwordspacing

\bibitem{bailey_survey_2009}
\BIBentryALTinterwordspacing
M.~Bailey, E.~Cooke, F.~Jahanian, Y.~Xu, and M.~Karir,
  ``\BIBforeignlanguage{en}{A {Survey} of {Botnet} {Technology} and
  {Defenses}},'' in \emph{\BIBforeignlanguage{en}{2009 {Cybersecurity}
  {Applications} \& {Technology} {Conference} for {Homeland}
  {Security}}}.\hskip 1em plus 0.5em minus 0.4em\relax Washington, DC, USA:
  IEEE, Mar. 2009, pp. 299--304. [Online]. Available:
  \url{http://ieeexplore.ieee.org/document/4804459/}
\BIBentrySTDinterwordspacing

\bibitem{roesch_snort_1999}
M.~Roesch, ``Snort - lightweight intrusion detection for networks,'' in
  \emph{Proceedings of the 13th USENIX Conference on System Administration},
  ser. LISA '99.\hskip 1em plus 0.5em minus 0.4em\relax USA: USENIX
  Association, 1999, p. 229–238.

\bibitem{wetherall_privacy_2011}
D.~Wetherall, D.~Choffnes, B.~Greenstein, S.~Han, P.~Hornyack, J.~Jung,
  S.~Schechter, and X.~Wang, ``Privacy revelations for web and mobile apps,''
  in \emph{Proceedings of the 13th {USENIX} conference on {Hot} topics in
  operating systems}, ser. {HotOS}'13.\hskip 1em plus 0.5em minus 0.4em\relax
  USA: USENIX Association, May 2011, p.~21.

\bibitem{le_pochat_tranco_2019}
\BIBentryALTinterwordspacing
V.~Le~Pochat, T.~Van~Goethem, S.~Tajalizadehkhoob, M.~Korczynski, and
  W.~Joosen, ``\BIBforeignlanguage{en}{Tranco: {A} {Research}-{Oriented} {Top}
  {Sites} {Ranking} {Hardened} {Against} {Manipulation}},'' in
  \emph{\BIBforeignlanguage{en}{Proceedings 2019 {Network} and {Distributed}
  {System} {Security} {Symposium}}}.\hskip 1em plus 0.5em minus 0.4em\relax San
  Diego, CA: Internet Society, 2019. [Online]. Available:
  \url{https://www.ndss-symposium.org/wp-content/uploads/2019/02/ndss2019_01B-3_LePochat_paper.pdf}
\BIBentrySTDinterwordspacing

\bibitem{public_suffix_list}
\BIBentryALTinterwordspacing
``Public {Suffix} {List}.'' [Online]. Available:
  \url{https://publicsuffix.org/}
\BIBentrySTDinterwordspacing

\bibitem{dietrich_investigating_2018}
\BIBentryALTinterwordspacing
C.~Dietrich, K.~Krombholz, K.~Borgolte, and T.~Fiebig,
  ``\BIBforeignlanguage{en}{Investigating {System} {Operators}' {Perspective}
  on {Security} {Misconfigurations}},'' in
  \emph{\BIBforeignlanguage{en}{Proceedings of the 2018 {ACM} {SIGSAC}
  {Conference} on {Computer} and {Communications} {Security}}}.\hskip 1em plus
  0.5em minus 0.4em\relax Toronto Canada: ACM, Oct. 2018, pp. 1272--1289.
  [Online]. Available: \url{https://dl.acm.org/doi/10.1145/3243734.3243794}
\BIBentrySTDinterwordspacing

\bibitem{aws_cloudformation}
\BIBentryALTinterwordspacing
``{AWS CloudFormation}.'' [Online]. Available:
  \url{https://aws.amazon.com/cloudformation/}
\BIBentrySTDinterwordspacing

\bibitem{terraform}
\BIBentryALTinterwordspacing
``{Terraform}.'' [Online]. Available: \url{https://www.terraform.io/}
\BIBentrySTDinterwordspacing

\bibitem{noauthor_best_2019}
\BIBentryALTinterwordspacing
``\BIBforeignlanguage{en}{Best {Practices} for {Setting} {Static} {IP}
  {Addresses} on {Cisco} {Business} {Hardware}},'' Dec. 2019. [Online].
  Available:
  \url{https://www.cisco.com/c/en/us/support/docs/smb/General/Best-practices-for-setting-a-static-IP-addresses-in-Cisco-Small-Business.html}
\BIBentrySTDinterwordspacing

\bibitem{noauthor_introducing_2021}
\BIBentryALTinterwordspacing
``\BIBforeignlanguage{en-US}{Introducing {IPv6}-only subnets and {EC2}
  instances},'' Nov. 2021, section: Amazon VPC. [Online]. Available:
  \url{https://aws.amazon.com/blogs/networking-and-content-delivery/introducing-ipv6-only-subnets-and-ec2-instances/}
\BIBentrySTDinterwordspacing

\bibitem{oltrogge_pin_2015}
\BIBentryALTinterwordspacing
M.~Oltrogge, Y.~Acar, S.~Dechand, M.~Smith, and S.~Fahl,
  ``\BIBforeignlanguage{en}{To {Pin} or {Not} to {Pin}—{Helping} {App}
  {Developers} {Bullet} {Proof} {Their} {TLS} {Connections}},'' 2015, pp.
  239--254. [Online]. Available:
  \url{https://www.usenix.org/conference/usenixsecurity15/technical-sessions/presentation/oltrogge}
\BIBentrySTDinterwordspacing

\bibitem{noauthor_ipv6_nodate}
\BIBentryALTinterwordspacing
``\BIBforeignlanguage{en-US}{{IPv6} {Adoption} in 2021}.'' [Online]. Available:
  \url{https://labs.ripe.net/author/stephen_strowes/ipv6-adoption-in-2021/}
\BIBentrySTDinterwordspacing

\bibitem{amazon_vpc}
\BIBentryALTinterwordspacing
``\BIBforeignlanguage{en-US}{Amazon {Virtual} {Private} {Cloud} ({VPC})}.''
  [Online]. Available: \url{https://aws.amazon.com/vpc/}
\BIBentrySTDinterwordspacing

\end{thebibliography}

\section*{Acknowledgement}

The authors would like to thank Michael Bailey for insightful discussion on our measurement techniques. We also thank the anonymous reviewers for productive feedback throughout the review process. This material is based upon work supported by the National Science Foundation Graduate Research Fellowship Program under Grant No.  DGE1255832. This work was supported in part by the NSF grant: CNS-1900873. Any opinions, findings, and conclusions or recommendations expressed in this material are those of the author(s) and do not necessarily reflect the views of the National Science Foundation.

\appendices
\section{Vulnerability Disclosures}
\label{appendix:disclosure}

As one might expect, the disclosures required by this study were complex. The number of organizations involved rendered it logistically impossible for our academic research to perform tenant disclosures. We divided our disclosure into two phases, one working with Amazon and the second through a series of direct tenant organization communications (email followed by a virtual discussion). We discuss the timing and processes associated with each of these activities below.

\subsection{Amazon-Coordinated Disclosure}

We first became aware of the scale of the results presented throughout in late May 2021. Following internal discussion and study validation, it became clear that direct handling of the disclosures by us would be logistically impossible. Moreover, we observed that the cloud provider (in this case Amazon AWS) would be in a better position to aid affected organizations, identify other vulnerable tenants, and augment their internal infrastructure to detect and prevent exploitation in the future (see Amazon mitigations in \autoref{sec:mitigations}). We reached out to Amazon May 27, 2021 as an initial contact. We have met with the AWS technical team weekly since June 4, 2021 and continue as of this submission for review.

The substance of these meetings initially supported disclosures, and evolved into broader discussions of causes and mitigations. During our first meeting we shared the identified tenant vulnerabilities and characterized the scope and detail of our results. We provided (a) a set of PowerPoint slides detailing the scope and method of this study, (b) a list of the affected domains, IP addresses and timestamps, (c) a list of servers provisioned by experimental apparatus, and (d) a list of vulnerable cloud configurations detected (AWS identifiers associated with cloud services such SNS topics, CloudFront distribution IDs). Beyond this, we have not shared {\it any} other captured tenant-specific data such as received headers, PCAP traces or other artifacts. We continue to share new results, detection methods, and mitigations as they are identified with AWS as discoveries warrant.

Amazon has informally shared results of their internal audits and discussed their plans for tenant notification (expected to occur in late August 2021). The bidirectional conversations have evolved to topics of root causes and mitigations, including those adopted by Amazon and discussed in the previous sections. Note that all statements and commentary attributed to Amazon in this paper have been confirmed for accuracy by their technical staff prior to submission.

\subsection{Direct Tenant Disclosure}\hypertarget{mr3b}{}

The second class of disclosures were performed by directly contacting tenants (see \autoref{results:root_causes}). The 17 organizations were initially contacted in mid-July 2021 and discussions continued until first week of August. All discussions were online and lasted from \SI{20}{} minutes to an hour, depending on the organization and the data we were disclosing.

The process for each discussion was the same. We would initially reach out via email to the security office or other contact within the organization indicating that we had found an issue as part of our study (which we would provide a brief overview of in the email). An online meeting was scheduled. Members of our research team and staff, as would a number of personnel from the organization. The call would begin with a brief overview of the study, its high level results, and then provide a brief discussion of the data we had relating to the organization. Lastly, we would introduce the questions identified in \refappendix{appendix:questions}. We proposed that they could answer the questions in the meeting, via email response, or not at all.

Immediately after the online meeting we would follow up with an email containing (a) a set of PowerPoint slides detailing the scope and method of this study, (b) metadata on affected organizational servers that received traffic (IP, timestamp created, length of time server ran), (c) the related PCAPs of traffic received, and (d) log data of requests (timestamp, domain, header data), and (e) the list of questions relating to root causes. Organizations would reply to these emails with responses at their discretion.

\section{Data Handling}\label{appendix:data_handling}

Sensitive data was received during our experiment, including tenant organizational data, credentials, and PII. For this reason, we created the following process for handling the data:

\begin{enumerate}
    
    \item All collected data was stored in an encrypted S3 bucket (AWS storage unit).
    
    \item Collected data was migrated to a special purpose storage/compute server. Storage was fully encrypted and the server was only used to store and process study data.
    
    \item Only one person in the team was given credentials to the server. Physical access to the server was also restricted.
    
    \item Any sharing of the study data (\eg, for the purpose of sharing with tenants following disclosure) was performed using secure transfer protocols and limited to the minimal necessary data.\hypertarget{or2a}{}
    
    \item Data will be securely deleted when it is no longer needed. The device will be securely wiped at the conclusion of this study.
    
\end{enumerate}

\noindent
Note that the data handling process was documented (in much more detail), reviewed, and approved by the security team (CISO Office) of our home institution, and covered under our IRB exemption.

\section{Other Disclosures}

While our measurement study was performed and found concrete vulnerabilities on Amazon Web Services, the nature of our findings suggests that the same issues are likely prevalent on other public clouds. As such, disclosures to other major cloud providers are ongoing as this paper is being submitted.

\section{Disclosure Questions}
\label{appendix:questions}
After sending disclosures to organizations, we performed disclosure discussions to better understand the scope and impact of Cloud Squatting attacks. For transparency, we detail our scripted questions below:

\begin{enumerate}
    \item Was the organization aware that domains pointed to stale cloud IP addresses?
    \item What organizations within your organization would be involved in the mitigation of the leakage and the potential notification process?
    \item What kinds of services is your organization currently deploying in the cloud?
    \item Are these services commercial or internally developed? To the degree that your organization can share, what kinds of technical platforms are they developed on, \ie, middleware stacks, programming languages, cloud configuration management and/or third-party software? 
    \item For services deployed, are these internal or externally facing, \ie, for internal use of customer interaction?
    \item Is the organization aware or can the organization speculate on the root causes of this vulnerability?
    \item Please add any additional information or comments the organization feels are relevant.
\end{enumerate}

\noindent
Note that we indicated in our meetings and emails that we committed to anonymizing the data returned and would not directly quote any response without prior approval.

\clearpage

\section{Traffic Analysis Architecture}\label{appendix:funnel}

Our filtering proceeds in four steps, roughly mirroring the network, transport, session, and application layers of the OSI model. Here we provide extended description of several of these steps.

\subsubsection{Transport Filtering}

\begin{figure}[H]
    \centering
        \centering
        \includegraphics[width=\columnwidth]{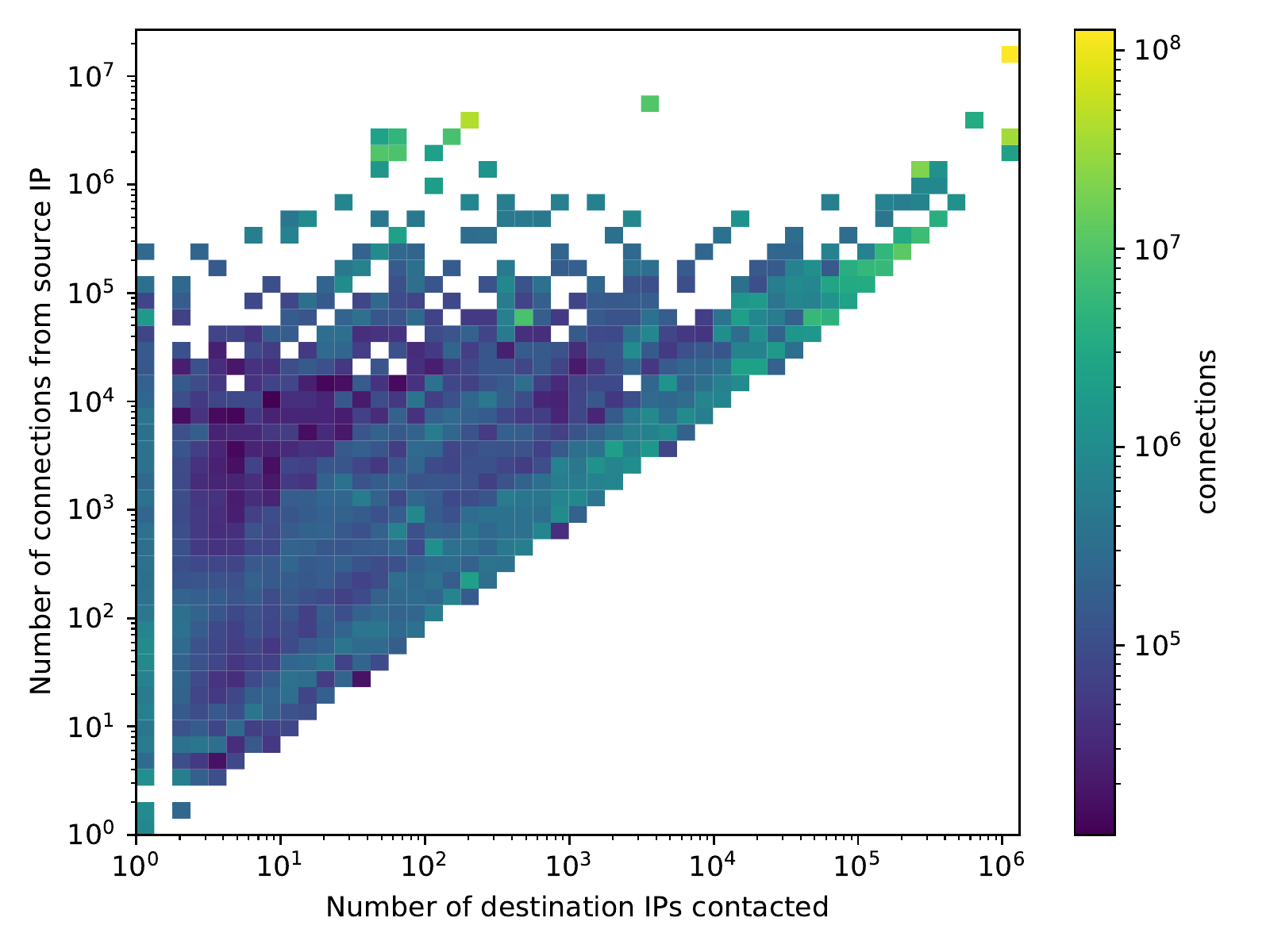}
    \caption{Distribution of IPs contacted and total connections from each source IP. Source IPs with a large number of connections also generally connect to a large number of IPs}
    \label{fig:ip_conn_dist}
\end{figure}

In the transport step, we consider the IPs and ports on our servers accessed by each remote IP. Intuitively, a client that accesses many IP addresses or many ports is likely scanner or exploit traffic. To this end, we collect statistics on the number of (IP, port) pairs contacted by each remote IP, and then filter our dataset to only remote IPs that access a single IP or port. Note that, while this may filter out legitimate exploitable traffic, this is an acceptable trade-off to reduce the size of the filtered dataset.

\autoref{fig:no_dups} shows the distributions of number of IPs and ports contacted and \autoref{fig:ip_conn_dist} shows the distribution of number of connections from IPs. We note similar access patterns here to those seen by ~\cite{richter_scanning_2019} on a popular content delivery network, though there are some key differences: (1) our figures demonstrate that session density is largely uniform across the distribution, (2) sessions attributed to large senders (\ie, those initiating many sessions) made a substantial contribution to the total sessions considered.

\subsubsection{Session Filtering}

As many web scanners are implemented using distributed botnets~\cite{antonakakis_understanding_2017, bailey_survey_2009}, checking for duplicate IP/port access does not effectively filter all traffic. To target less sophisticated scanners (such as Mirai~\cite{antonakakis_understanding_2017}) that are distributed, we analyze and filter based on TCP session behavior. Two types of traffic are removed: (1) traffic that simply sends \texttt{SYN} packets to scan available ports, and (2) traffic that completes TCP session initialization but does not send any payload. In the first case, this traffic is clearly not exploitable because no connection is established. In the second, this traffic may be legitimate or exploitable, but uses a protocol (such as Telnet) where the server is expected to initiate the protocol. Such protocols are not as readily studied by our approach and can therefore be excluded. This step has no false negatives with respect to our detection goals, as sessions that do not have a client payload would yield no interesting information during manual analysis.

\subsubsection{Application Filtering}

We finally filter traffic at the application layer. This step is the most complex and time-intensive, so the previous filter steps contribute to the step being tractable. In this step, we use existing network intrusion detection rules from Snort~\cite{roesch_snort_1999}, as well as additional manually-generated rules targeted at peer-to-peer and other cloud traffic:

\begin{itemize}
    \item \textit{Exploits}. Snort is designed to detect exploit traffic against specific vulnerabilities, but much of the traffic we measured targets configuration errors with shellcode exploits. To this end we exclude traffic with markers of very common shellcode exploits: \texttt{wget}, \texttt{curl}, \texttt{chmod}, \texttt{curl}, \texttt{shell}. We also exclude \texttt{dnp3} and \texttt{tds} protocols as manual analysis showed these legacy protocols were exclusively shellcode spam.
    \item \textit{Peer-to-peer}. Because the bootstrapping protocols of many peer-to-peer protocols involve random IP scans, these protocols made a large contribution to uninteresting traffic. Further, the semantics of these protocols make them uninteresting for exploitation as they place no trust in the connected server and do not rely on potentially-latent configuration. We filtered Bittorrent, Bitcoin, Skype, and IPFS traffic.
    \item \textit{Proxy traffic}. We filtered large amounts of HTTP traffic relating to proxy connections, likely intended to exploit or research other network security issues. These filters removed HTTP \texttt{CONNECT} methods and requests where the request Path contained a full URI (i.e., starting with http://). Such behavior is not generally used for proxy traffic.
    \item \textit{Health checks}. We received many HTTP Health Check requests from Amazon Route53. Because this traffic is studied independently it was excluded from this analysis.
\end{itemize}

\end{document}